\pdfoutput=1
\documentclass[usenatbib]{mn2e}
\usepackage[T1]{fontenc}
\usepackage[latin9]{inputenc}
\usepackage{array}
\usepackage{amsmath}
\usepackage{graphicx}
\newcommand{\apj}{ApJ}
\newcommand{\apjs}{ApJS}
\newcommand{\apjl}{ApJL}

\newcommand{\aap}{A~\&~A}
\newcommand{\aapr}{Astron.~\&~Astrophys.~Rev.}
\newcommand{\mnras}{MNRAS}

\newcommand{\physrep}{Phys. Rep.}
\newcommand{\araa}{An. Rev. Astron. Astroph.}
\newcommand{\ssr}{Space Sci. Rev.}

\providecommand{\tabularnewline}{\\}

\usepackage{bm}
\usepackage{amsbsy}
\usepackage{amstext}

\title[Nonlinear regimes in mean-field full-sphere dynamo]{Nonlinear regimes in mean-field full-sphere dynamo}

\author[V.V.~Pipin]
{V.V. Pipin \thanks{email: pip@iszf.irk.ru}\\
$^{1}${Institute of Solar-Terrestrial Physics, Russian Academy of Sciences,
Irkutsk, 664033, Russia}}
\begin{document}
\maketitle
\begin{abstract}
{The mean-field dynamo model is employed to study the non-linear dynamo
regimes in a fully convective star of mass 0.3$M_{\odot}$ rotating
with period of 10 days. For
the intermediate value of the  parameter of the turbulent magnetic Prandl number,
$Pm_{T}=3$ we found the oscillating dynamo regimes with period about
40Yr. The higher $Pm_{T}$ results to longer dynamo periods.  If the large-scale
flows is fixed we find that the dynamo transits from axisymmetric
to non-axisymmetric regimes for the overcritical parameter of the
$\alpha$effect. The change of dynamo regime occurs because of the
non-axisymmetric non-linear $\alpha$-effect. The situation persists
in the fully non-linear dynamo models with regards of the magnetic
feedback on the angular momentum balance and the heat transport in
the star. It is found that the large-scale magnetic field quenches
the latitudinal shear in the bulk of the star. However, the strong
radial shear operates in the subsurface layer of the star. In the
nonlinear case the profile of the angular velocity inside the star
become close to the spherical surfaces. This supports the equator-ward
migration of the axisymmetric magnetic field dynamo waves. It was found that, the magnetic configuration of
the star dominates by the regular non-axisymmetric mode m=1, forming
Yin Yang magnetic polarity pattern with the strong (>500 G) poloidal
magnetic field in polar regions.}
\end{abstract}
\begin{keywords}
Activity - stars:  magnetic field - stars: dynamo
\end{keywords}

\section{Introduction}

Stars with the extended convective envelopes demonstrate the high
level of magnetic activity \citep{mdwa,2009ARAA_donat,L1-2015SSRv}.
It is commonly believed that the magnetic activity of these stars
origins from the hydromagnetic turbulent dynamo action \citep{brsu05,Brunrev14}.
Extremely high magnetic activity was found on the fully-convective
low-mass stars which belong to the M-dwarfs branch of the low main
sequence of the Hertzsprung-Rawssel diagram. Observations of the M-dwarfs
indicated the rather strong large-scale magnetic field with strength
of several kG \citep{S1-1985ApJ,S2-1986ApJ,J1-1996ApJ,L1-2015SSRv}.
The magnetic topology of the M-dwarfs is likely depends on the mass
and the rotation period of a star \citep{2009ARAA_donat,2016MNRAS1129S}.
Observations indicate that the early type M-dwarfs with the moderate
period of rotation about 4-5 days demonstrate the strong non-axisymmetric
magnetic field with the dominant toroidal component \citet{D2-2008MNRAS}.
The extremely fast rotating early type M-dwarfs with period of rotation
less than 1 day indicate the transition to the axisymmetric dynamo
with the dominant poloidal component of the large-scale magnetic field.
Situation become complicated on the mid and late-type M-stars which
have the masses less than $0.2M_{\odot}$ as they could show either
the strong axisymmetric dipole-kind large-scale magnetic field, or
the low-strength non-axisymmetric magnetic field \citep{M1=2008MNRAS,M2-2010MNRAS}.
Thus we can conclude about three basic states of the dynamo on the
fast rotating M-dwarfs, they are: the strong multipole magnetic field
(hereafter, SM), the strong dipole field (hereafter SD) and the weak
multipole (hereafter WM) magnetic field. We follow notation suggested
by \citet{2011MNRAS.418L.133M}. Interesting that simultaneously with
multiply states of the dynamo regimes, the dynamo generated total
magnetic flux do not show the rotation-activity connection which is
known among the solar type stars \citep{2003ApJ583.451M}.

Observed magnetic properties of the M-stars initiated the number of
the theoretical studies employing the mean-field models (see, \citealt{2006AA446.1027C,elst07,KMS14,shul15})
and the direct numerical simulations (e.g., \citealt{2006ApJ...638..336D,2008ApJ676.1262B,2013IAUS294.163D,2014AA564A.78S}).
Using the results of the numerical simulations, \citet{2011MNRAS.418L.133M}
suggests that the bi-stability of the magnetic topology on the late-type
M-stars could result from two types of the convection regimes occurred
in the fast rotating convective bodies \citep{1988GApFD44.3R}. Also,
the direct numerical simulations show the differential rotation is
important part of the dynamo in the fully convective stars. Similar
conclusions were suggested by \citet{shul15} after studying the linear
dynamo regimes. 

Current interpretation of the dynamo bi-stability given by \citet{2011MNRAS.418L.133M}
suggests that the strength of the large-scale magnetic field is compatible
with the nonlinear balance between the Lorentz and Coriolis force
in case of the SD-type magnetism and it is established by the Lorentz-inertia
force balance in case of the WM magnetism. Later, \citet{2014AA564A.78S}
found that in the anelastic simulations the separation between the
SD and WM magnetism is less profound than they as well as others (e.g.,
\citealt{2009EL8519001S}) found with the Boussinescue approximation.
The origin of the strong multipolar magnetic field on the moderate
rotating early M-stars is barely studied. Results of the mean-field
models and the numerical simulations suggest the dynamo on these stars
could operate with help of the differential rotation. The linear analysis
of \citet{KMS14} show that the axisymmetric magnetic field modes
have the smaller critical threshold of the dynamo instability than
the non-axisymmetric ones. Thus, the transition from axisymmetric
to non-axisymmetric dynamo can occur only in the nonlinear regime. 

The paper we study the nonlinear dynamo models for a fully convective
star. Here we restrict ourselves to the same case of the star discussed
earlier by \citet{shul15}, i.e., the star of mass 0.3$M_{\odot}$
of 1Gyr age and rotating with period of 10 days. We will address the
axisymmetric and non-axisymmetric dynamo regimes with regards for
the non-linear back reaction of the large-scale magnetic field on
the $\alpha$-effect and the large-scale flow. The solution of the
dynamo problem is coupled with the solution of the mean angular momentum
balance and the mean heat transport in the convective sphere. The
main goal of the paper is to find the typical topology of the large-scale
magnetic field in the nonlinear dynamo for the given rotation period
and investigate the nonlinear effects on the dynamo.

\section{Model formulation}

We consider a fully convective star of mass 0.3$M_{\odot}$ of 1Gyr
age and rotating with period of 10 days. The reference internal thermodynamic
structure of the star was calculated using the MESA stellar evolution
code, the version r7503, \citep{mesa11,mesa13}. It is assumed that
composition of the star is similar to the Sun and the metallicity
parameter is $Z=0.02$. In the reference model we neglect effects
of stellar rotation on the hydrostatic equilibrium. The convection
parameters in the MESA code are determined by $\alpha_{MLT}={\displaystyle \frac{\ell}{H_{p}}}=1.9$,
where $H_{p}$is the pressure stratification scale. For the given
parameters the star has the radius of $R_{\star}\approx0.286R_{\odot}$,
the luminosity of $L_{\star}\approx0.1354L_{\odot}$ and the surface
temperature of $3520$K. Current understanding of the internal structure
of the fully convective stars is not complete. That's why the theoretical
predictions for the stellar radius and the $T_{eff}$ of the M-dwarfs
are different from observation\citep{mdwa}.

\subsection{Heat transport and angular momentum balance\label{sub:Heat-transporT}}

The mean-field heat transport equation takes into account effects
of the global rotation on the thermal equilibrium. It is calculated
from the mean-field heat transport equation, 
\begin{equation}
\overline{\rho}\overline{T}\frac{\partial\overline{s}}{\partial t}+\overline{\rho}\overline{T}\left(\overline{\mathbf{U}}\cdot\boldsymbol{\nabla}\right)\overline{s}=-\boldsymbol{\nabla}\cdot\left(\mathbf{F}^{conv}+\mathbf{F}^{rad}\right)+\epsilon,\label{eq:heat}
\end{equation}
where $\epsilon$ is the source function, $\mathbf{\overline{U}}$
is axisymmetric mean flow, $\overline{\rho}$ and $\overline{T}$
are the mean density and temperature, and $\overline{s}$ is the mean
entropy. In what follows, the over-bar denote the axisymmetric component
of the mean field and the angle brackets are used for the ensemble
average of the field which could contain the large-scale non-axisymmetric
modes contributions as well.

We employ expression of the anisotropic convective flux suggested
by \citet{1994AN....315..157K} (hereafter KPR94), 
\begin{equation}
F_{i}^{conv}=-\overline{\rho}\overline{T}\chi_{ij}\nabla_{j}\overline{s},\label{conv}
\end{equation}
where in the heat eddy-conductivity tensor $\chi_{ij}$ we have to
take into account both the global rotation and the large-scale magnetic
field effects. The expression has the complicated form and it is unknown
for the general case if both the Coriolis and the Lorenz forces are
not small simultaneously. This corresponds to conditions in the considered
M-star. The fast rotation regime and $\Omega^{*}>1$, holds in whole
volume of the star except the layer above $r=0.975R_{\star}$ (see,
Fig.1a), where the Coriolis number $\Omega^{*}=2\tau_{c}\Omega_{\star}$,
and $\tau_{c}$ is the turn-over time of convective flow. In the paper
we approximate the eddy heat conductivity tensor in following to \citet{phd}:
\begin{equation}
\chi_{ij}=\chi_{T}\left(\phi_{\chi}^{(I)}\left(\beta\right)\phi\left(\Omega^{*}\right)\delta_{ij}+\phi_{\chi}^{(A)}\left(\beta\right)\phi_{\parallel}\left(\Omega^{*}\right)\frac{\Omega_{i}\Omega_{j}}{\Omega^{2}}\right).\label{eq:ht-f}
\end{equation}
The effect of the global rotation on the heat transport depends on
and the functions $\phi$ and $\phi_{\parallel}$ are defined in KPR4.
The magnetic feedback on the eddy heat-conductivity depends on the
functions $\phi_{\chi}^{(I)}$ and $\phi_{\chi}^{(A)}$,(see Appendix)
and the parameter $\beta={\displaystyle \frac{\left|B\right|}{\sqrt{4\pi\overline{\rho}}u'^{2}}}$,
where $\left|B\right|$ is the strength of the large-scale magnetic
field, and $u'$ is the RMS convective velocity. Note that for the
case $\beta>1$ we have $\phi_{\chi}^{(I)}\sim\beta^{-2}$ and $\phi_{\chi}^{(A)}\sim\beta^{-1}$.
Thus the isotropic part of the eddy heat conductivity is quenched
stronger than that in direction of the rotation axis. However in the
case of the weak magnetic field the expression returns to the case
discussed in KPR94.

The self-consistent model could also include effects the Joule's heating
or sinks of the convective energy into magnetic activity, see e.g.,
\citet{1992AA...265..328B} and \citet{2004ARep...48..418P}. We put
off discussion of those effects. In limits of the slow rotation and
the weak magnetic field, i.e., $\Omega^{*}\rightarrow0$, and $\beta\rightarrow0,$
the heat conductivity tensor reduces to isotropic form, $\chi_{ij}={\displaystyle \frac{1}{3}\delta_{ij}\ell}u'$,
where $\ell$ is the mixing length. The RMS convective velocity is
determined from the mixing-length relations 
\[
u'=\frac{\ell}{2}\sqrt{-\frac{g}{c_{p}}\frac{\partial\overline{s}}{\partial r}}.
\]

The integration domain of the mean-field model is from $r_{i}=0.05R_{\star}$
to $r_{e}=0.98R_{\star}$, we exclude the central and the near-surface
regions. At the inner boundary the total flux $F_{r}^{conv}+F_{r}^{rad}={\displaystyle \frac{L_{\star}\left(r_{i}\right)}{4\pi r_{i}^{2}}}$
and for the external boundary, in following to \citet{kit11}, we
use 
\[
F_{r}=\frac{L_{\star}}{4\pi r_{e}^{2}}\left(1+\left(\frac{s}{c_{p}}\right)^{4}\right).
\]
We put other details about the mean-field model of the heat transport
in Appendix.

The heat transport equation is coupled to equations of the angular
momentum balance and the mean-field dynamo equations. . In the spherical
coordinate system the conservation of the angular momentum \citep{1989drsc.book.....R}
it expressed as follows: 
\begin{equation}
\frac{\partial}{\partial t}\overline{\rho}r^{2}\!\!\sin^{2}\!\!\theta\Omega\!=\!-\!\boldsymbol{\nabla\!\!\cdot\!\!}\left(\!\!r\sin\theta\left(\!\!\overline{\rho}\hat{\mathbf{T}}_{\phi}\!\!+\!\!r\overline{\rho}\sin\theta\Omega\mathbf{\overline{U}^{m}\!\!}-\!\!\frac{\left\langle \mathbf{B}\right\rangle \left\langle B_{\phi}\right\rangle }{4\pi}\!\!\right)\!\!\right),\label{eq:az}
\end{equation}
where, $\left\langle \mathbf{B}\right\rangle $ is the large-scale
dynamo generated magnetic field (see, the Subsection 2.2). The mean
flow satisfies the continuity equation, 
\begin{equation}
\boldsymbol{\nabla}\cdot\overline{\rho}\mathbf{\overline{U}}=0,\label{eq:cont}
\end{equation}
where $\mathbf{\overline{U}}=\mathbf{\overline{U}}^{m}+r\sin\theta\Omega\hat{\mathbf{\boldsymbol{\phi}}}$
and $\boldsymbol{\hat{\phi}}$ is the unit vector in the azimuthal
direction. The equation for the azimuthal component of the large-scale
vorticity , $\omega=\left(\boldsymbol{\nabla}\times\overline{\mathbf{U}}^{m}\right)_{\phi}$
, is 
\begin{eqnarray}
\frac{\partial\omega}{\partial t}\!\!\! & \!\!=\!\!\!\! & r\sin\theta\boldsymbol{\nabla}\cdot\left(\frac{\hat{\boldsymbol{\phi}}\times\boldsymbol{\nabla\cdot}\overline{\rho}\hat{\mathbf{T}}}{r\overline{\rho}\sin\theta}-\frac{\mathbf{\overline{U}}^{m}\omega}{r\sin\theta}\right)\!\!+r\sin\theta\frac{\partial\Omega^{2}}{\partial z}\label{eq:vort}\\
 & +\!\!\! & \frac{1}{\overline{\rho}^{2}}\left[\boldsymbol{\nabla}\overline{\rho}\times\boldsymbol{\nabla}\overline{p}\right]_{\phi}\!\!\nonumber \\
 & + & \!\frac{1}{\overline{\rho}^{2}}\left[\!\!\boldsymbol{\nabla}\overline{\rho}\times\left(\!\!\boldsymbol{\nabla}\frac{\left\langle \mathbf{B}\right\rangle ^{2}}{8\pi}-\frac{\left(\left\langle \mathbf{B}\right\rangle \boldsymbol{\cdot\nabla}\right)\left\langle \mathbf{B}\right\rangle }{4\pi}\!\right)\!\!\right]_{\phi}\nonumber 
\end{eqnarray}
where $\boldsymbol{\hat{\phi}}$ is a unit vector in azimuthal direction,
$\hat{\mathbf{T}}$ is the turbulent part of the Reynolds and Maxwell
stresses and $\partial/\partial z=\cos\theta\partial/\partial r-\sin\theta/r\cdot\partial/\partial\theta$
is the gradient along the axis of rotation. The turbulent stresses
include the non-disspative part due to the $\Lambda$-effect and the
anisotropic eddy viscosity. The theory is not complete because it
does not comprise the joint effect of the global rotation and the
large-scale magnetic field on the angular momentum transport. We apply
the theory developed in \citet{kuetal96} and \citet{phd} for the
case of the arbitrary $\Omega^{*}$ and the arbitrary strength of
the large scale magnetic field. Theirs derivations are valid in the
case when the toroidal component of the large-scale magnetic field
dominates the poloidal one. In equation for the toroidal vorticity,
Eq.(\ref{eq:vort}) we neglect the radial derivative of the Lorentz
force in compare to the density gradient. The details about implementation
of the turbulent stress tensor $\hat{\mathbf{T}}$ are given in Appendix. 

\begin{figure*}
\includegraphics[width=1\textwidth]{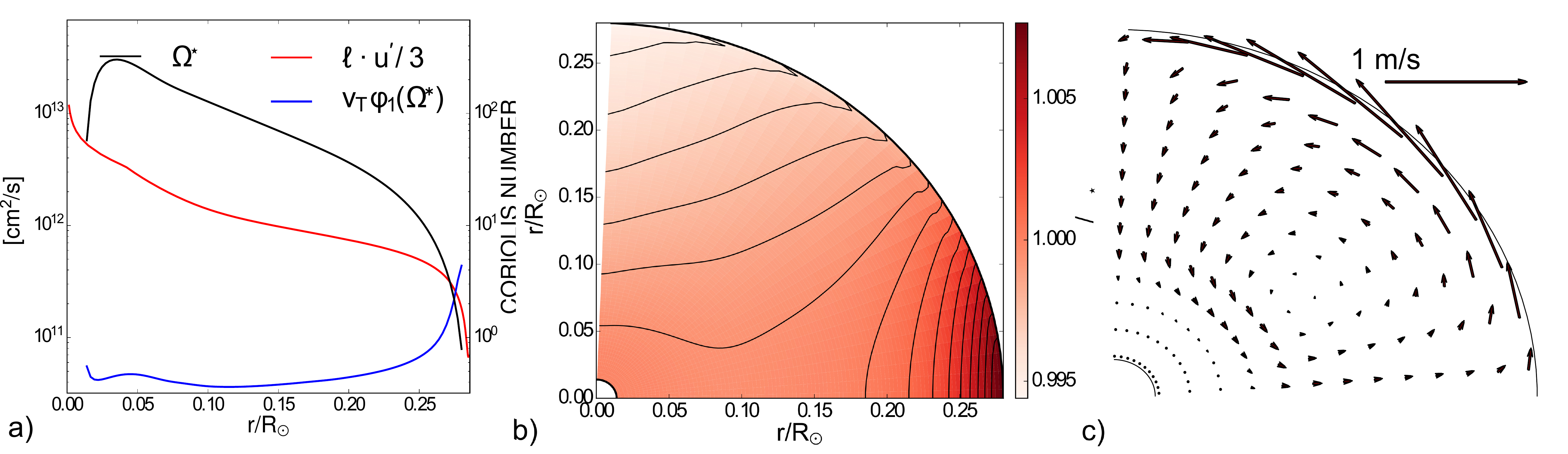} \caption{\label{fig:basic}a) The Coriolis number $\Omega^{*}=2\tau_{c}\Omega_{\star}$
(black line), where $\tau_{c}$ is the turn-over time of convection
(from the MESA code), the turbulent diffusivity parameter, red line;
the blue line show isotropic eddy viscosity from the heat transport
model; b) angular velocity profiles with contour levels which cover
the range of values depicted on the color bar; c) geometry of the
meridional circulation, in the Northern hemisphere.}
\end{figure*}

Figure (\ref{fig:basic}) shows profiles of the internal parameters
of the mean-field model of the heat transport and the angular momentum
balance together with some input parameters from the MESA code. It
is found that the convective turnover time varies from about of 1
day at the near-surface layer to about of 200 days near the center
of the star. This results to the strong modifications of the turbulent
viscosity parameters. The resulted differential rotation is rather
weak, it is about of 0.01$\Omega_{\star}$, where $\Omega_{\star}=7.25\cdot10^{-6}$rad/s
is the stellar rotation rate. The given angular velocity profile corresponds
qualitatively to results of \citet{KMS14}. The angular velocity profile
shows conical isolines pattern in the bulk of the star. This pattern
changes to the cylinder like pattern in the equatorial region. In
the portion of the star which occupied by the weakly varying angular
velocity, the given pattern is different to results of the direct
numerical simulations (cf, \citealt{2008ApJ676.1262B,2015ApJ813L31Y}).
However, the strong shear in equatorial region is presented in all
models. The meridional circulation consists of the one cell in each
hemisphere with poleward flow in upper part of the star. The amplitude
of the flow is about one meter per second at the surface.

\subsection{Dynamo model}

The dynamo model takes into account both the axisymmetric and the
non-axisymmetric large-scale magnetic field. Its evolution is described
by the mean-field induction equation \citep{KR80}: 
\begin{equation}
\partial_{t}\left\langle \mathbf{B}\right\rangle =\boldsymbol{\nabla}\times\left(\boldsymbol{\mathcal{E}}+\left\langle \mathbf{U}\right\rangle \times\left\langle \mathbf{B}\right\rangle \right)\label{eq:mfe}
\end{equation}
where $\boldsymbol{\mathcal{E}}=\left\langle \mathbf{u\times b}\right\rangle $
is the mean electromotive force; $\mathbf{u}$ and $\mathbf{b}$ are
the turbulent fluctuating velocity and magnetic field respectively;
and $\left\langle \mathbf{U}\right\rangle $ and $\left\langle \mathbf{B}\right\rangle $
are the mean velocity and magnetic field. We remind that in the paper,
the angle brackets are used for the ensemble average of the field
which could contain the large-scale non-axisymmetric modes contributions
as well as the axisymmetric modes of the mean field which is denoted
by the over-bar. The mean flow is axisymmetric, i.e., $\left\langle \mathbf{U}\right\rangle \equiv\overline{\mathbf{U}}$,
and it is determined from solution of the angular momentum balance.
In the fully nonlinear case the solution of the angular momentum balance
is coupled with the mean-field dynamo equations and the mean-field
equation for the heat transport.

Let $\hat{\boldsymbol{\phi}}=\mathbf{e_{\phi}}$ and $\hat{\mathbf{r}}=r\mathbf{e}_{r}$
be vectors in the azimuthal and radial directions respectively, then
we represent the mean magnetic field vectors as follows: 
\begin{eqnarray}
\left\langle \mathbf{B}\right\rangle  & = & \overline{\mathbf{B}}+\tilde{\mathbf{B}}\label{eq:b0}\\
\mathbf{\overline{B}} & = & \hat{\boldsymbol{\phi}}B+\nabla\times\left(A\hat{\boldsymbol{\phi}}\right)\label{eq:b1}\\
\tilde{\mathbf{B}} & = & \boldsymbol{\nabla}\times\left(\hat{\mathbf{r}}T\right)+\boldsymbol{\nabla}\times\boldsymbol{\nabla}\times\left(\hat{\mathbf{r}}S\right),\label{eq:b2}
\end{eqnarray}
where $\overline{\mathbf{B}}$ is the axisymmetric, and $\tilde{\mathbf{B}}$
is non-axisymmetric part of the large-scale magnetic field, $A$,
$B$, $T$ and $S$ are scalar functions. Hereafter, the non-axisymmetric
part of the mean field is denoted by the wave above the symbol.

We employ the mean electromotive force in form: 
\begin{equation}
\mathcal{E}_{i}=\left(\alpha_{ij}+\gamma_{ij}\right)\left\langle B\right\rangle _{j}-\eta_{ijk}\nabla_{j}\left\langle B\right\rangle _{k}.\label{eq:EMF-1}
\end{equation}
where symmetric tensor $\alpha_{ij}$ models the generation of magnetic
field by the $\alpha$- effect; antisymmetric tensor\textbf{ }$\gamma_{ij}$
controls the mean drift of the large-scale magnetic fields in turbulent
medium, including the magnetic buoyancy; tensor $\eta_{ijk}$ governs
the turbulent diffusion. Some details about the $\boldsymbol{\mathcal{E}}$
are given in appendix, (also, see, \citealt{pi08Gafd}).

In our model the $\alpha$ effect takes into account the kinetic and
magnetic helicities in the following form: 
\begin{eqnarray}
\alpha_{ij} & = & C_{\alpha}\psi_{\alpha}(\beta)\alpha_{ij}^{(H)}\eta_{T}+\alpha_{ij}^{(M)}\frac{\left\langle \chi\right\rangle \tau_{c}}{4\pi\overline{\rho}\ell^{2}}\label{alp2d}
\end{eqnarray}
where $C_{\alpha}$ is a free parameter which controls the strength
of the $\alpha$- effect due to turbulent kinetic helicity; $\alpha_{ij}^{(H)}$
and $\alpha_{ij}^{(M)}$ express the kinetic and magnetic helicity
parts of the $\alpha$-effect, respectively; $\eta_{T}=\nu_{T}/Pm_{T}$
is the magnetic diffusion coefficient, $Pm_{T}$ is the turbulent
magnetic Prandtl number and $\left\langle \chi\right\rangle =\left\langle \mathbf{a}\cdot\mathbf{b}\right\rangle $
($\mathbf{a}$ and $\mathbf{b}$ are the fluctuating parts of magnetic
field vector-potential and magnetic field vector). Both the $\alpha_{ij}^{(H)}$
and $\alpha_{ij}^{(M)}$ depend on the Coriolis number. Function $\psi_{\alpha}(\beta)$
controls the so-called ``algebraic'' quenching of the $\alpha$-
effect where $\beta=\left\langle \left|\mathbf{B}\right|\right\rangle /\sqrt{4\pi\overline{\rho}u'^{2}}$,
$u'$ is the RMS of the convective velocity.

The magnetic helicity conservation results to the dynamical quenching
of the dynamo. Contribution of the magnetic helicity to the $\alpha$-effect
is expressed by the second term in Eq.(\ref{alp2d}). The magnetic
helicity density of turbulent field, $\left\langle \chi\right\rangle =\left\langle \mathbf{a}\cdot\mathbf{b}\right\rangle $,
is governed by the conservation law \citep{pip13M}: 
\begin{equation}
\frac{\partial\left\langle \chi\right\rangle ^{(tot)}}{\partial t}=-\frac{\left\langle \chi\right\rangle }{R_{m}\tau_{c}}-2\eta\left\langle \mathbf{B}\right\rangle \cdot\left\langle \mathbf{J}\right\rangle -\boldsymbol{\nabla\cdot\mathcal{F}}{}^{\chi},\label{eq:helcon-1}
\end{equation}
where $\left\langle \chi\right\rangle ^{(tot)}=\left\langle \chi\right\rangle +\left\langle \mathbf{A}\right\rangle \cdot\left\langle \mathbf{B}\right\rangle $
is the total magnetic helicity density of the mean and turbulent fields,
$\boldsymbol{\mathcal{F}}^{\chi}=-\eta_{\chi}\boldsymbol{\nabla}\left\langle \chi\right\rangle $
is the diffusive flux of the turbulent magnetic helicity, and $R_{m}$
is the magnetic Reynolds number. The coefficient of the turbulent
helicity diffusivity, $\eta_{\chi}$, is chosen ten times smaller
than the isotropic part of the magnetic diffusivity , $\eta_{\chi}=\frac{1}{10}\eta_{T}$.
The magnetic helicity conservation is determined by the magnetic Reynolds
number $R_{m}$. In this paper we employ $R_{m}=10^{4}$.

The numerical scheme employs the spherical harmonics decomposition
for the non-axisymmetric part of the problem. At the bottom of the
domain we put the potentials $S$ and $T$ , as well as the axisymmetric
fields, $B$ and $A$ to zero. At the top the poloidal field is smoothly
matched to the external potential field and the toroidal field goes
to zero

The numerical scheme employs the pseudo-spectral approach for integration
along latitude and the finite differences along the radius. Fort the
non-axisymmetric part of the problem we employ the spherical harmonics
decomposition, i.e., the scalar functions $T$ and $S$ are represented
in the form: 
\begin{eqnarray}
T\left(r,\mu,\phi,t\right) & = & \sum\hat{T}_{l,m}\left(r,t\right)\bar{P}_{l}^{\left|m\right|}\exp\left(im\phi\right),\label{eq:tdec}\\
S\left(r,\mu,\phi,t\right) & = & \sum\hat{S}_{l,m}\left(r,t\right)\bar{P}_{l}^{\left|m\right|}\exp\left(im\phi\right),\label{eq:sdec}
\end{eqnarray}
where $\bar{P}_{l}^{m}$ is the normalized associated Legendre function
of degree $l\ge1$ and order $m\ge1$. The simulations which we will
discuss include 310 spherical harmonics (up to $l_{max}=20$). Note
that $\hat{S}_{l,-m}=\hat{S}_{l,m}^{*}$ and the same for $\hat{T}$.
All the nonlinear terms are treated explicitly in the real space.
The numerical integration is carried out in latitude from the pole
to pole and in radius from $r_{b}=0.05R_{\star}$ to $r_{e}=0.98R_{\star}$.

\begin{figure}
\includegraphics[width=1\columnwidth]{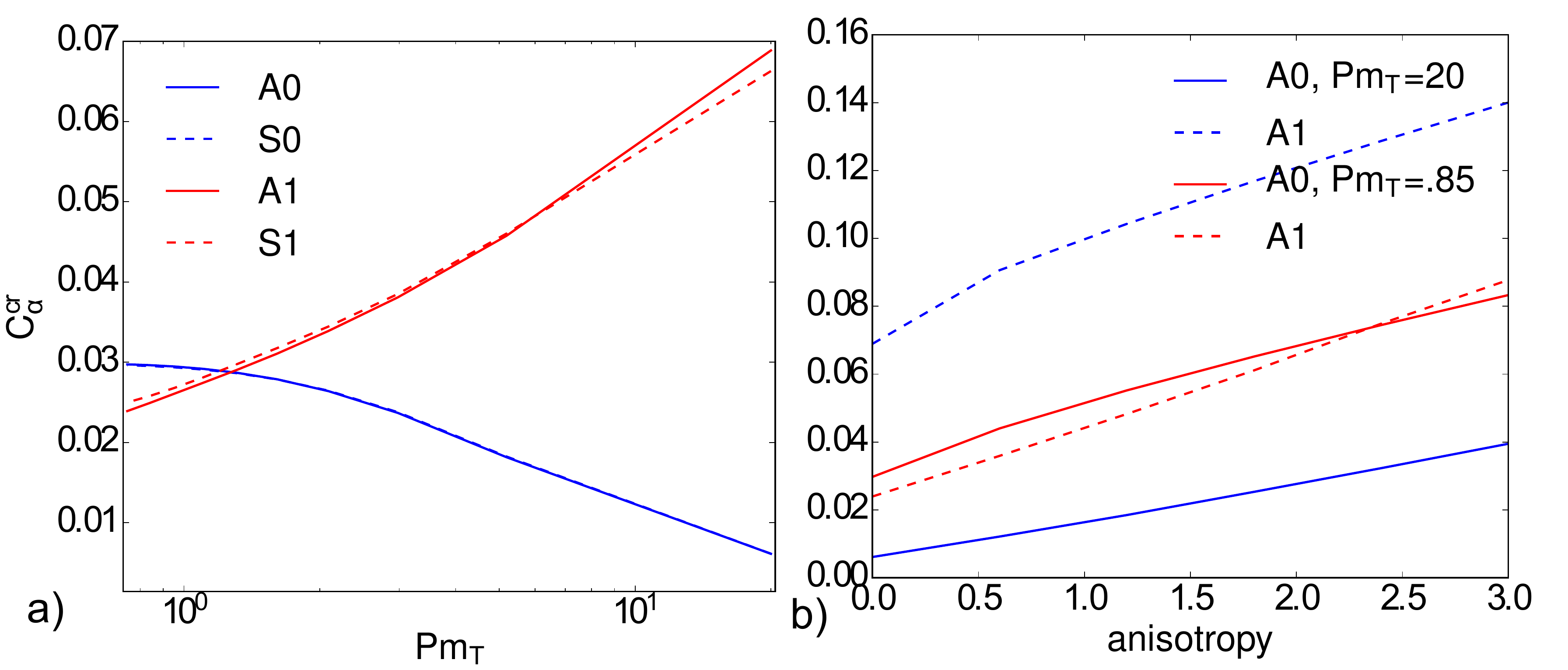} \caption{\label{fig-lin}The critical threshold parameter $C_{\alpha}^{(cr)}$
for isotropic diffusivity, $A=0$; b) The dependence of the critical
threshold parameter $C_{\alpha}^{(cr)}$ on the anisotropy of the
turbulent diffusivity, for $Pm_{T}=20$ (blue lines), and $Pm_{T}=0.85$,
(red lines).}
\end{figure}

The thermal equilibrium, the angular momentum balance and evolution
of the large-scale magnetic field is controlled by the free parameters,
which are the angular velocity of the global rotation $\Omega_{0}=7.25\times10^{-6}{\rm rad/s}$,
the turbulent Prandtl number $Pr_{T}={\displaystyle \frac{\nu_{T}}{\chi_{T}}}$,
the turbulent magnetic Prandtl number $Pm{}_{T}={\displaystyle \frac{\nu_{T}}{\eta_{T}}}$
, the parameter of the magnetic field generation by the $\alpha$-effect,
$C_{\alpha}$, and the magnetic Reynolds number $R_{m}$. We use the
mixing-length expression for the eddy heat conductivity, ${\displaystyle \chi_{T}=\frac{\ell^{2}}{6}\sqrt{-\frac{g}{c_{p}}\frac{\partial\overline{s}}{\partial r}}}.$
In the all models we fix $Pr_{T}={\displaystyle {\displaystyle \frac{3}{4}}}$,
$Pm{}_{T}=3$, and $R_{m}=10^{4}$. We will discuss the possible dependence
of results on $Pm_{T}$ , as well.

We studied the eigenvalue dynamo problem before running the nonlinear
models. In the linear model we neglect the radial dependence of the
$\alpha$-effect and turbulent diffusivity. Solutions of the eigenvalue
problem showed that the linear properties of the dynamo model are
in agreement with results reported earlier by \citet{elst07} and
\citet{shul15}. More specifically, the results of the linear problem
solutions are as follows. Firstly, for the high $Pm_{T}$ the axisymmetric
dynamo has smaller the critical dynamo instability threshold instability
than the non-axisymmetric dynamo. The transition from axisymmetric
to non-axisymmetric regimes occurs for $Pm_{T}\approx1$. This is
in agreement with findings of \citet{shul15}. Also the solution shows
that the critical threshold for the symmetric and antisymmetric about
equator dynamo modes are close and the symmetric modes have the smaller
threshold than the antisymmetric ones. Secondly, it was found that
for the case $Pm_{T}\approx1$, when the non-axisymmetric dynamo instability
is more powerful than the axisymmetric one, the rotationally induced
anisotropy of the magnetic diffusivity can promotes the dynamo instability
of the axisymmetric magnetic field if the amplitude of the eddy diffusivity
along the rotation axis is twice of that one in the perpendicular
direction. This result is in agreement with that reported by \citet{elst07}.
Figures \ref{fig-lin}(a,b) illustrate our findings. In comparing
our results with findings from reported in above cited papers we have
to take into account that the parameter of the $\alpha$-effect in
the reduced linear models contain the density stratification factor,
$\boldsymbol{\tilde{\Lambda}}^{(\rho)}=R_{\odot}\boldsymbol{\nabla}\log\overline{\rho}$,
and its mean value in the star is $\overline{\left|\boldsymbol{\tilde{\Lambda}}^{(\rho)}\right|}\approx50$.
Also models of \citet{shul15} were normalized for diffusivity $10^{11}\mathrm{cm^{2}/s}$
and it is $10^{13}\mathrm{cm^{2}/s}$ in our model.

\begin{table*}
\caption{\label{tab:nonlinear}Basic parameters, $B_{max}$ is the maximum
of the magnetic field strength in the star, $\overline{B}_{\|}$ and
$\tilde{B}_{\|}$ are the\emph{ mean} strength of the mean poloidal
components of the axisymmetric and non-axisymmetric magnetic field
at the surface, $\overline{B}_{\perp}$ and $\tilde{B}_{\perp}$ is
the same for the toroidal magnetic field below surface at the ${\displaystyle \frac{3}{4}}R_{\star}$,
M is the ratio of the energy of the non-axisymmetric magnetic field
to the total magnetic energy at the same radial distance and ${\displaystyle \overline{\frac{\Delta\Omega}{\Omega}}}$
is the measure of the latitudinal shear at the surface.}

\begin{tabular}{cc>{\centering}p{1.8cm}>{\centering}p{1cm}>{\centering}p{1.5cm}>{\centering}p{1.5cm}>{\centering}p{1.5cm}>{\centering}p{1cm}}
\hline 
 & $C_{\alpha}$  & Angular Momentum & $B_{max}$, {[}kG{]} & $\overline{B}_{\|}$, $\tilde{B}_{\|}$ {[}kG{]} & $\overline{B}_{\perp}$,$\tilde{B}_{\perp}$ {[}kG{]} & $M={\displaystyle \frac{\tilde{E}_{m}}{E_{m}}}$ & ${\displaystyle \overline{\frac{\Delta\Omega}{\Omega}}}$\tabularnewline
\hline 
M1  & 0.04  & no & 3 & 0.3, 0  & 1.5, 0 & 0 & 0.014\tabularnewline
\hline 
M2  & 0.04  & no & 4 & 0.2, 0.35 & 1, 2 & 0.5 & 0.014\tabularnewline
\hline 
M3 & 0.05 & no & 8 & 0.04, 0.8 & 0.2, 4 & 0.9 & 0.014\tabularnewline
\hline 
M4 & 0.04 & yes & 1.5 & 0.2 & 0.35 & 0 & 0.012\tabularnewline
\hline 
M5 & 0.05 & yes & 1.8 & 0.2,0.3 & 0.7,1 & 0.6 & 0.009\tabularnewline
\hline 
\end{tabular}
\end{table*}

\section{Results}

Table \ref{tab:nonlinear} contains general parameters of our models.
They are: the $B_{max}$ is the maximum strength of the large-scale
magnetic field in the star, the $\overline{B}_{\|}$ and $\tilde{B}_{\|}$
are the mean strength of the axisymmetric and non-axisymmetric large-scale
poloidal magnetic field on the surface, the $\overline{B}_{\perp}$
and $\tilde{B}_{\perp}$ is the same for the toroidal magnetic field
at the radial distance ${\displaystyle \frac{3}{4}}R_{\star}$, the
$M={\displaystyle \frac{\tilde{E}_{m}}{E_{m}}}$ is the ratio of the
energy of the non-axisymmetric mode of the large-scale magnetic field
to the total magnetic energy of the large-scale field at the radial
distance ${\displaystyle \frac{3}{4}}R_{\star}$, and the parameter
${\displaystyle \overline{\frac{\Delta\Omega}{\Omega}}}$ is the mean
latitudinal shear on the top of the integration domain, where the
mean is computed over one dynamo cycle. 

In the paper we show results for five different runs of the nonlinear
dynamo models. In all the runs we put $Pm_{T}=3$. In this case the
dynamo period is about 40 years. \citet{shul15} discussed linear
dynamo regimes with the longer dynamo period about 100Y. This is because
they employed the higher $Pm_{T}=10$ in thier models. We have nonlinear
runs with $Pm_{T}=10$ but not for all cases listed in the Table 1.
The higher $Pm_{T}$, the longer dynamo period and it takes longer
evolution time for the model to reach some stationary regime of the
dynamo oscillations, especially in the case of the non-axisymmetric
dynamo regimes. Also the effect of the meridional circulation for
the case of $Pm_{T}=10$ requires the better spatial resolution near
poles. We made separate runs for axisymmetric and non-axisymmetric
dynamo regimes. The latter models take into account both the axisymmetric
and non-axisymmetric magnetic field generation. In this paper we restrict
the study of the fully nonlinear model by the case of the $C_{\alpha}$
when both the axisymmetric and the non-axisymmetric magnetic fields
are unstable to generation in the large-scale dynamo. The initial
field in all runs has no preferable parity relative to the equator.

\begin{figure*}
\includegraphics[width=0.95\linewidth]{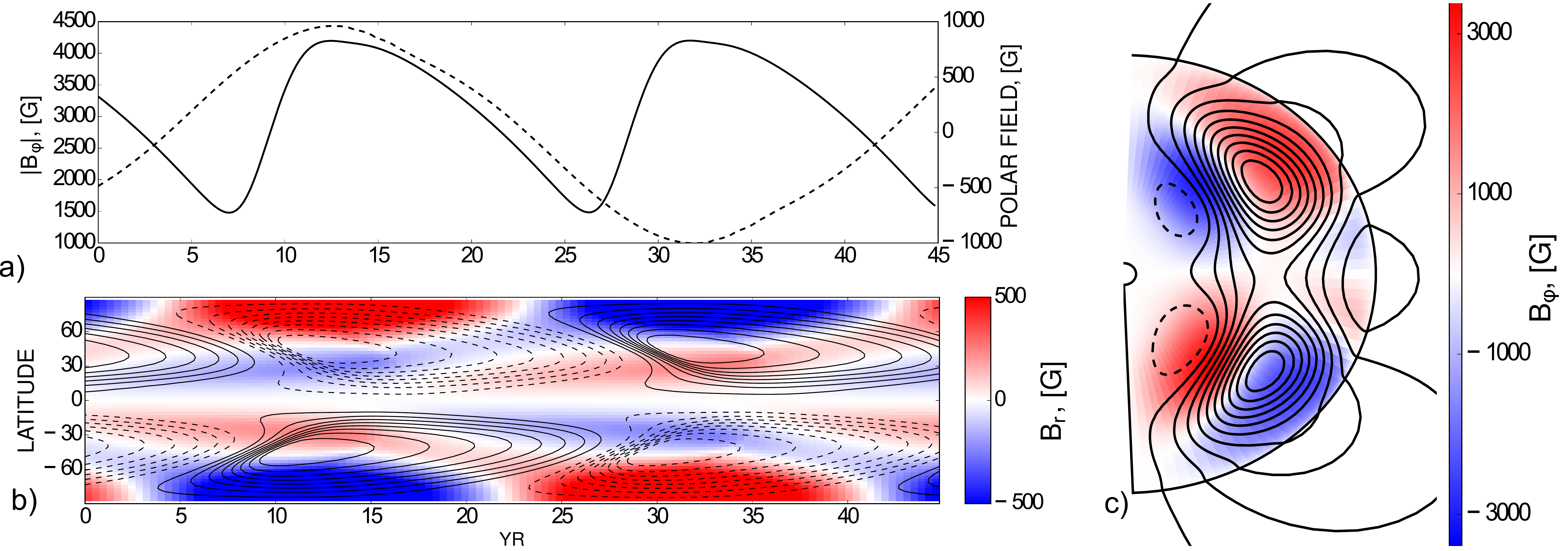}

\caption{\label{fig:M1l}The model M1, a) evolution of the mean strength toroidal
magnetic field at the $r=\frac{3}{4}R_{\star}$and the radial magnetic
field strength at the North pole (dashed line); b) the time-latitude
diagram for toroidal magnetic field (contours in range $\pm4$kG)
at the $r=\frac{3}{4}R_{\star}$, and the color image shows the radial
magnetic field at the surface; c) snapshot of the large-scale magnetic
field distributions at the growing phase of the cycle }
\end{figure*}

\begin{figure}
\includegraphics[width=0.95\linewidth]{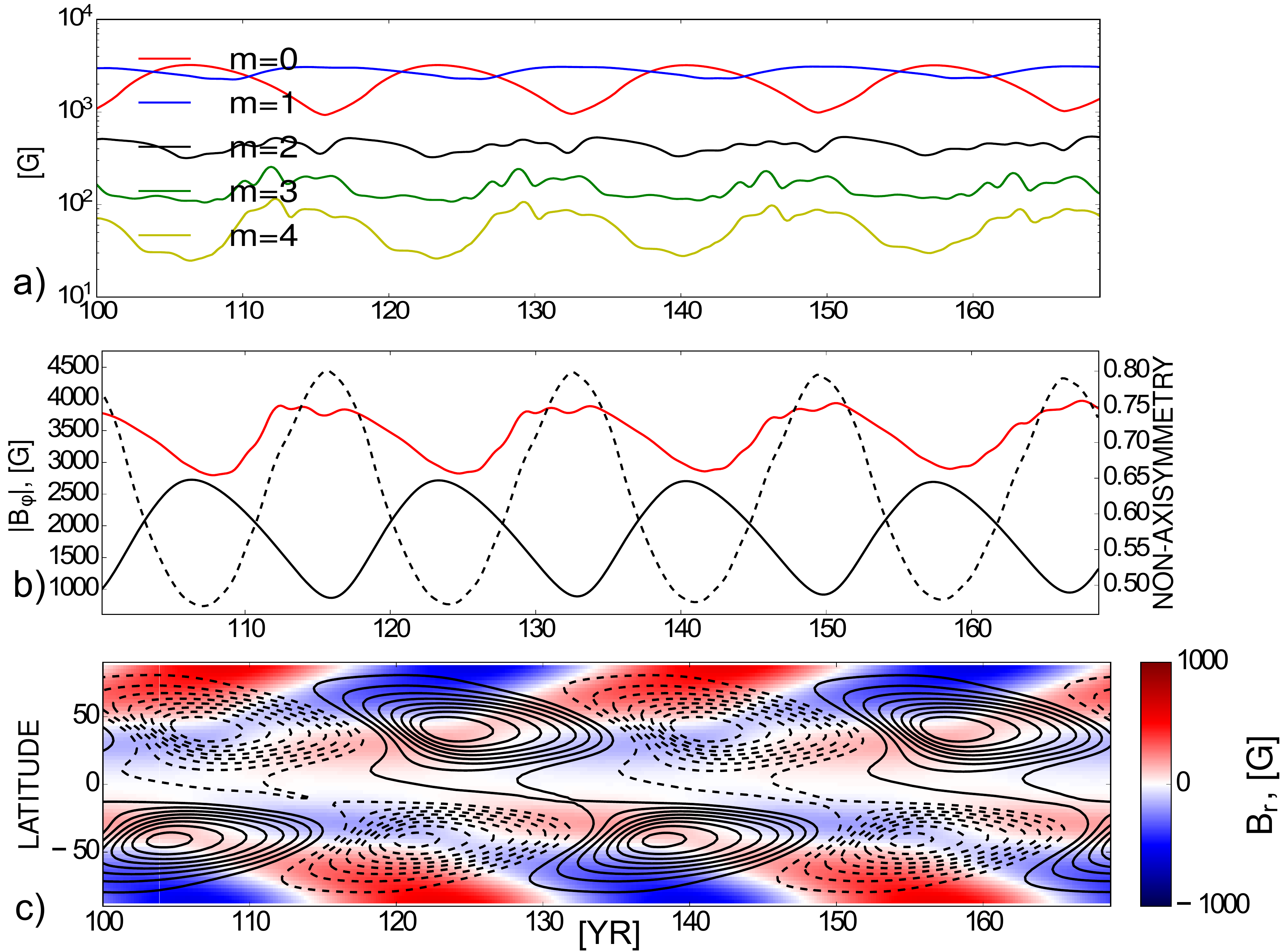}

\caption{\label{fig:M2l}The model M2, a) the mean strength of the first five
partial modes of the toroidal magnetic field at the $r=\frac{3}{4}R_{\star}$;
b) at the same $r$, the index of the non-axisymmetric of the large-scale
magnetic field (dashed line), the mean strength of the axisymmetric
toroidal magnetic field (solid line) and the mean strength of the
large-scale toroidal field taking into account both the axisymmetric
and non-axisymmetric parts of the magnetic field; c) the same as Fig.\ref{fig:M1l}b}
\end{figure}

\subsection{Nonlinear $\alpha$-effect}

In this subsection we consider the nonlinear models with magnetic
feedback on the generation of the large-scale magnetic fields by the
$\alpha$-effect. The models remain kinematic relative to the large-scale
flow. The non-linear $\alpha$-effect takes into account the dynamical
feedback due to magnetic helicity conservation (see, the Eq(\ref{eq:helcon-1}))
and the ``instantaneous'' quenching which is related with magnetic
feedback from the Lorentz forces on the turbulent convection. This
concept was originally formulated by \citet{kleruz82}. The Fig.\ref{fig:M1l}
show results for the model M1 which illustrates the axisymmetric dynamo
when the parameter of the parameter $C_{\alpha}$ is about factor
one and half of the critical dynamo instability threshold. The model
show the mixed parity solution with some preference to generation
of the antisymmetric about equator magnetic field. Butterfly diagrams
shows the solar-like equatorial drift of the toroidal magnetic field
of the strength 4kG at the $r=\frac{3}{4}R_{\star}$. The radial magnetic
field drift to the pole where it reaches strength of the 1kG during
the maximum of the dynamo cycle. The polar drift of the poloidal field
is supported by the meridional circulation. The migrating dynamo wave
of the poloidal magnetic field is transformed to the steady one when
the meridional circulation is neglected. Also, in this case the dominance
of the antisymmetric relative to equator magnetic field become less
clear in the nonlinear mixed parity solution. It is found that the
meridional flow has only a small effect on the amplitude of the dynamo
wave. 

\begin{figure}
\includegraphics[width=1\columnwidth]{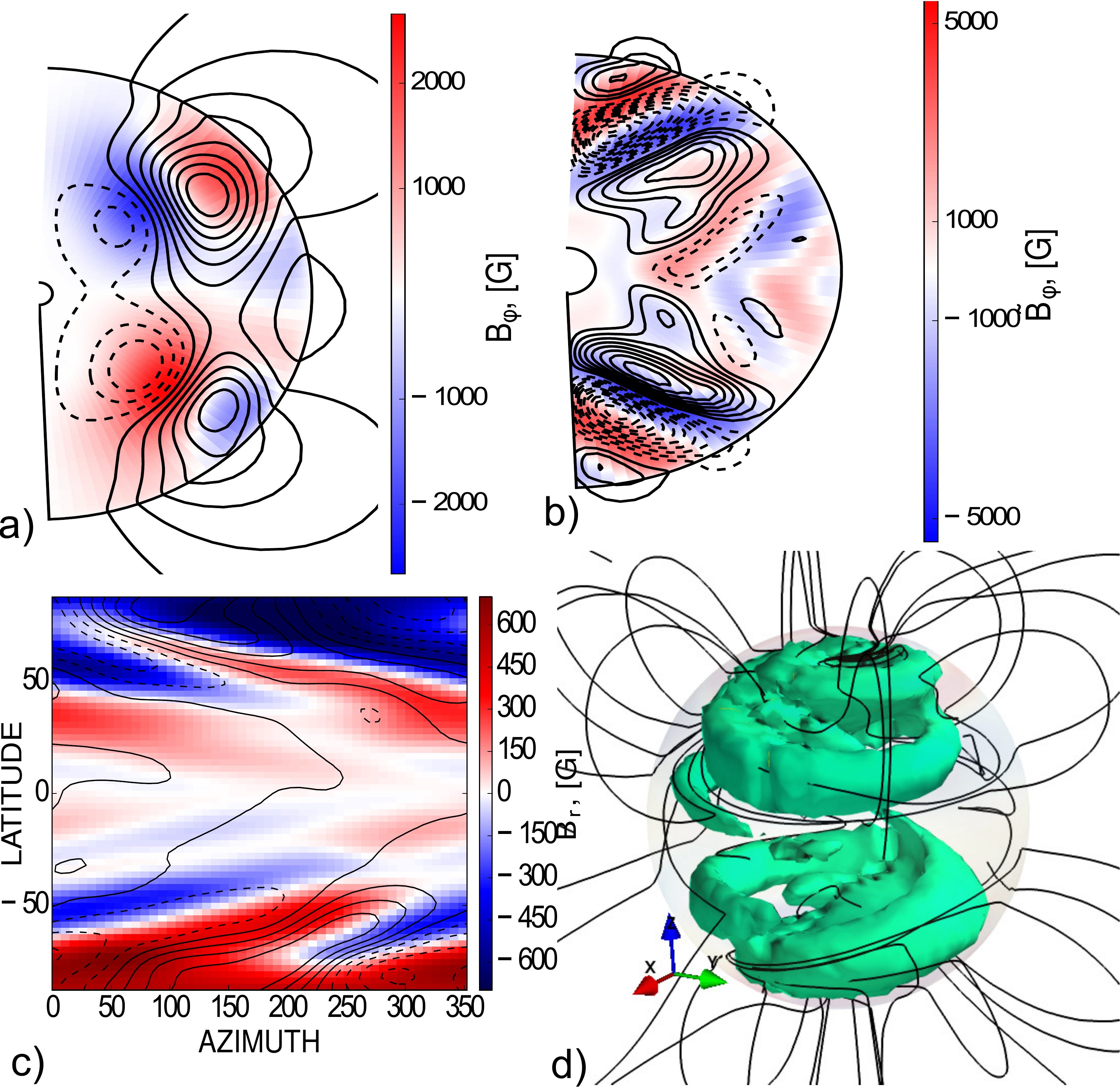}\caption{\label{fig:M2}The model M2, a) snapshot of the axisymmetric toroidal
magnetic field (color image) and the poloidal field at the end of
run ($t=170$ Yr); b) snapshot of distribution the non-axisymmetric
magnetic field at the longitude $\phi=0$ , color image shows the
toroidal magnetic field and contours - the poloidal magnetic field;
c) the non-axisymmetric radial magnetic field at the surface (color
image) and contours show the toroidal field at the subsurface layer;
d) magnetic field lines out of the star and surface show the large-scale
magnetic field of magnitude 1kG inside the star.}
\end{figure}

Results of the linear problem study show that the non-axisymmetric
magnetic field is unstable to generation for the same parameter $C_{\alpha}$
as it is employed in the model M1. The model M2 takes the non-axisymmetric
magnetic field into account. Fig.\ref{fig:M2l} show results for variations
of the magnetic energy, index of the non-axisymmetric and the time-latitude
diagrams of the axisymmetric magnetic field in the model. We find
that the non-axisymmetric dynamo quenches the strength of the generated
axisymmetric magnetic field. The most important quenching mechanisms
are due to effects of the magnetic helicity generation from the non-axisymmetric
dynamo and another effect is due to the magnetic buoyancy which is
increased when the magnetic energy increases. In our intepretation,
we have to take into account that the magnetic buoyancy can promote
the dynamo instability of the non-axisymmetric field \citep{2001ApJ551536D}. 

\begin{figure}
\includegraphics[width=1\columnwidth]{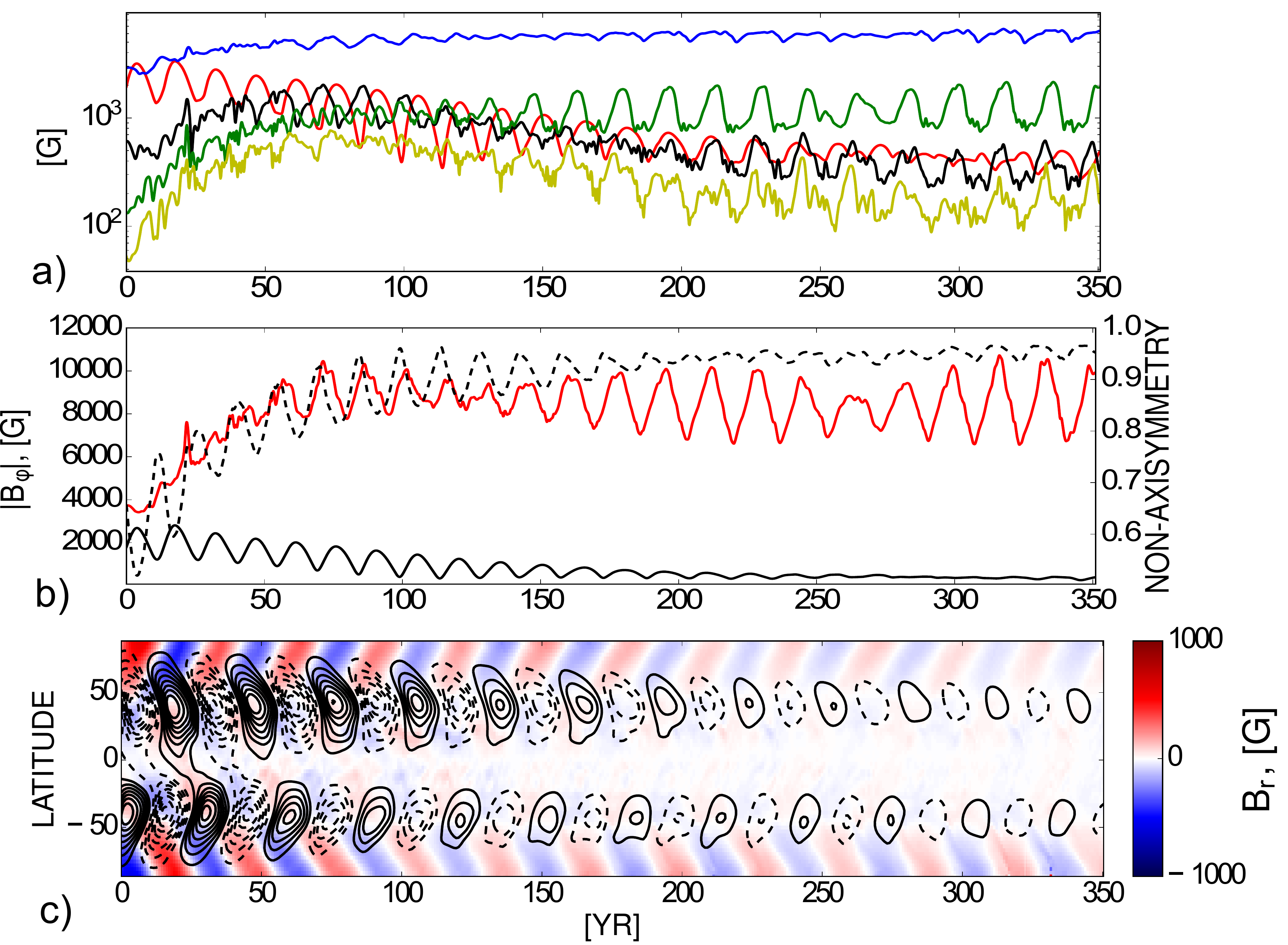}

\caption{\label{fig:M3l}The same as the Figure \ref{fig:M2l} for the model
M2.}
\end{figure}

The non-axisymmetric dynamo wave has a spiral pattern which is rigidly
rotating (see Fig \ref{fig:M2}d), which produces Yin Yang magnetic
polarity pattern on the surface of the star. The strength of the spiral
arms vary in time because of interaction with the axisymmetric magnetic
field. This causes the long-term variation of the magnetic energy
of the non-axisymmetric field at the given radial distance of the
star.

Figure \ref{fig:M2} show snapshots of the magnetic field distribution
inside and outside of the star. Configuration of the axisymmetric
field in the model M2 is similar that in the model M1. The non-axisymmetric
field is distributed along iso-surface of the angular velocity. The
polar regions on the surface of star are occupied by the mixture of
the axisymmetric radial magnetic and the non-axisymmetric field magnetic
field of the m=1 mode. 

\begin{figure}
\includegraphics[width=1\columnwidth]{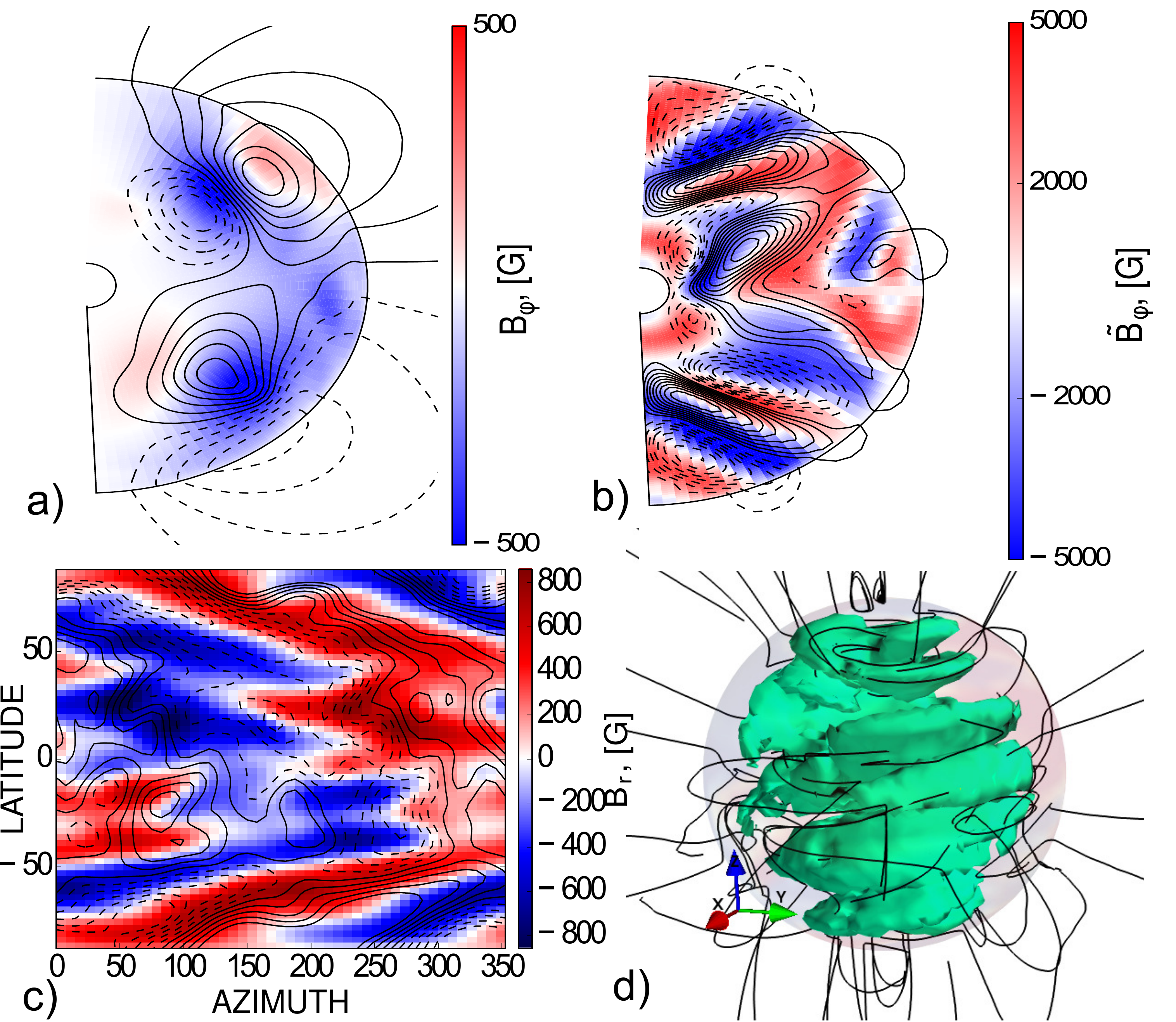}\caption{\label{fig:M3}The same as the Figure \ref{fig:M2} for the model
M3}
\end{figure}

In the model M3 the alpha-effect parameter $C_{\alpha}$ is about
twice of the dynamo instability threshold. The model M3 shows the
stronger quenching of the axisymmetric dynamo than the model M2. This
is illustrated in Figures \ref{fig:M3l} and \ref{fig:M3}. In fact,
the axisymmetric magnetic field nearly disappear in the stationary
phase of the evolution. At the surface the non-axisymmetric magnetic
field shows the large-scale spot-like pattern with angular size about
30$^{\circ}$. Those spots are located in the equatorrial and polar
regions as well.

\subsection{Fully non-linear dynamo}

This subsection contain results about the fully nonlinear dynamo with
regards for the magnetic feedback on the angular momentum balance
and heat transport inside the star. In the model M4 we neglect effects
of the non-axisymmetric field on the dynamo.

\begin{figure}
\includegraphics[width=1\columnwidth]{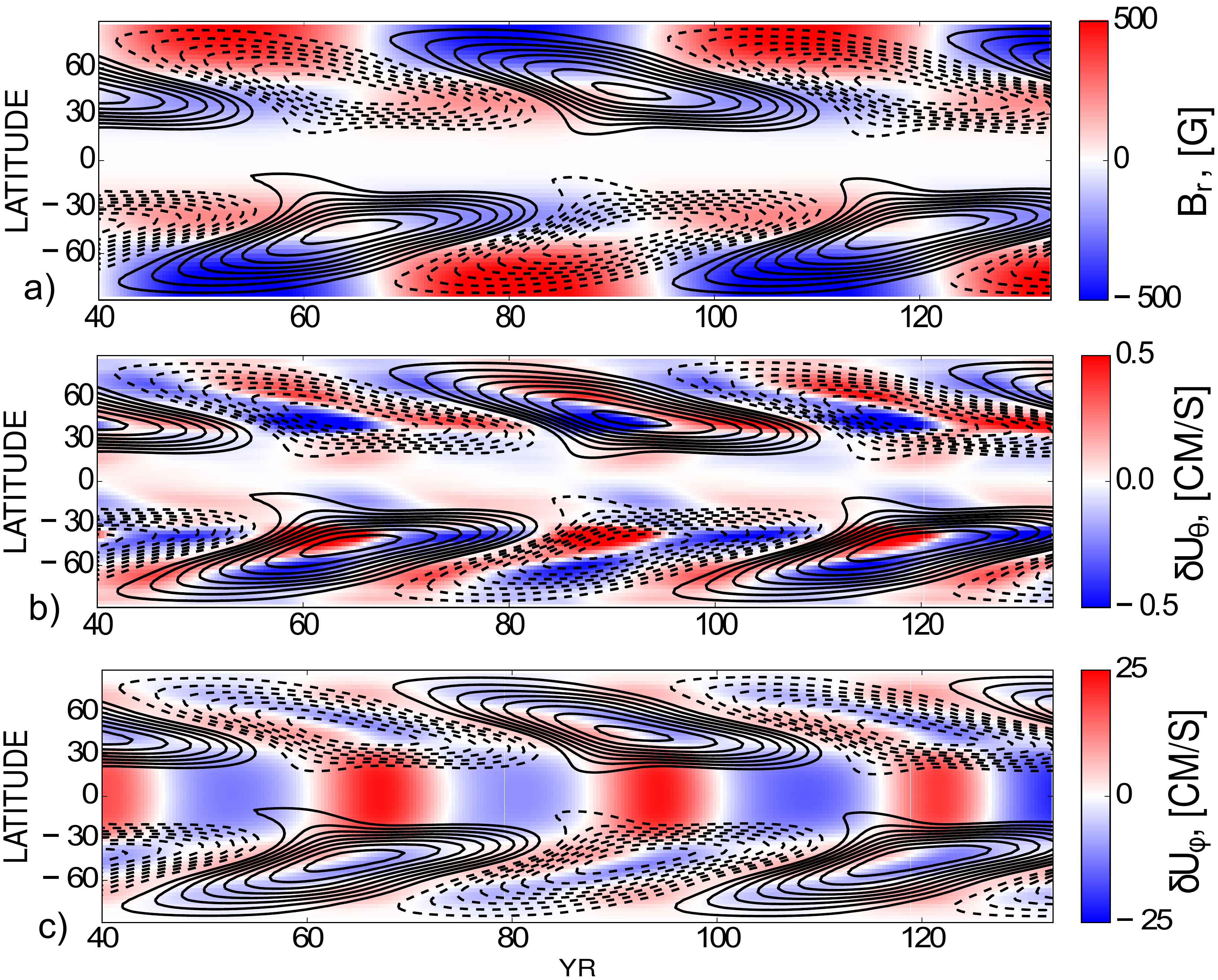}

\caption{\label{fig:M4l}The model M4, a) the time-latitude diagram for toroidal
magnetic field (contours in range $\pm1.5$kG) at the $r=\frac{3}{4}R_{\star}$,
and the color image shows the radial magnetic field at the surface;
b) color image show the variation of the latitudinal meridional flow
(positive to the equator) c) the same as b) for variation of the toroidal
velocity field at the surface;}
\end{figure}

Figure \ref{fig:M4l} show the time-latitude diagrams for variations
of the magnetic field, angular velocity and the latitudinal component
of the meridional flow in the model M4. The variations of the angular
velocity caused by the dynamo, can be observed as the azimuthal flow
waves. The model M4 demonstrate some similarity to the solar case,
i.e., the positive azimuthal flow wave is located on the equatorial
side of the dynamo wave of the large-scale toroidal field. Simultaneously,
the model shows the meridional flows direct to the maximum of the
toroidal magnetic field. Zonal variations of rotation is about 10
percents of the mean latitudinal shear. Variations of the meridional
circulation are smaller than one percent of the mean magnitude. The
surface mean latitudinal shear in model M4 is about 15 percents smaller
than in the kinematic models. Magnetic feedback on the differential
rotations reduces the strength of the toroidal field in the model
by factor 3 in compare to model M1. 

\begin{figure}
\includegraphics[width=1\columnwidth]{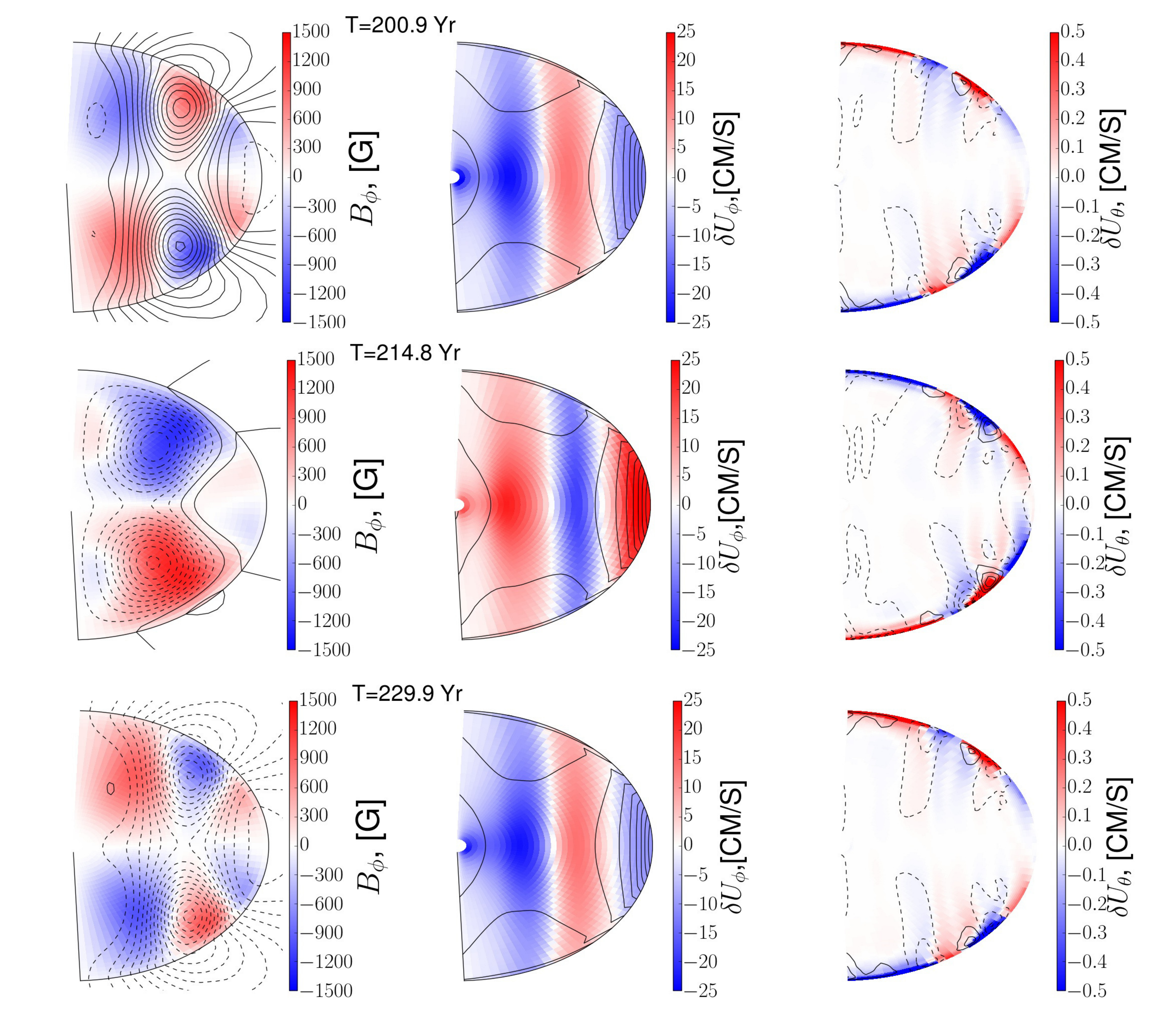}

\caption{\label{fig:M4_snaps}Model M4. Snapshots of the magnetic field (left),
angular velocity, the azimuthal zonal flow distributions (middle)
and the meridional circulation (right, color image is for the latitudinal
component and contours are for the radial component) for a half of
the dynamo cycle.}
\end{figure}

Figure \ref{fig:M4_snaps} show variations of the magnetic field and
angular velocity on period of half of the magnetic cycle. It is seen
that dynamo wave migrate outward of the rotation axis. The migration
of the dynamo waves induces variations of the angular velocity which
is separated to zones of the accelerated and decelerated motions.
Those zones are elongated along the rotation axis. The dynamo wave
inside star distorts distribution of the angular velocity bowing the
angular velocity iso-surface to equator. It makes the angular velocity
profile inside the star to become close to the spherical surfaces.
This supports the equatorial drift of the dynamo wave. In the upper
part of the star drift of the dynamo waves of the toroidal field follows
the distorted isolines of the angular velocity. The drift goes equator-ward
up to 30$^{\circ}$ latitude. The drifting wave of the poloidal field
is transformed to nearly steady on the surface because of effect of
the meridional circulation. We also see that rather small variations
of the meridional circulation are concentrated to the surface.

The reduced differential rotation results to a reduced ratio between
the mean strength of the toroidal and poloidal field. In the model
M4 it is about 2 and in the models M1, M2, M3 it is about 5. The strength
of the polar field in the model M4 is about 500 G which is by factor
2 smaller than in models M1 and M2.
\begin{figure*}
\includegraphics[width=1\textwidth]{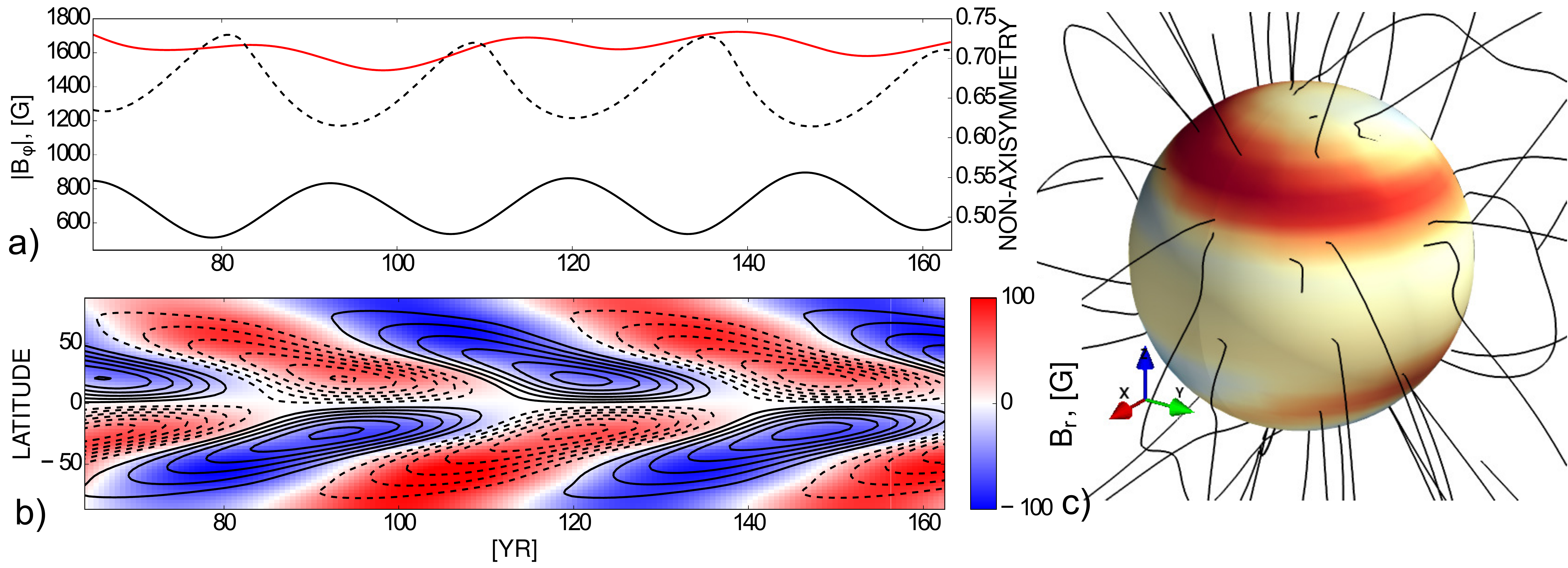}
\caption{\label{fig:M5e}The model M5, a) the mean strength of sum the first five
partial modes of the toroidal magnetic field at the $r=\frac{3}{4}R_{\star}$
(red line), the index of the non-axisymmetric of the large-scale
magnetic field (dashed line), the mean strength of the axisymmetric
toroidal magnetic field (solid black line); b) the time-latitude diagram
of the axisymmetric toroidal magnetic field at the $r=\frac{3}{4}R_{\star}$
(contours for the range $\pm500$G) and the axisymmetric radial magnetic
field at the surface; c) snapshot of the large-scale magnetic field
lines out of the star and the backgorund image shows the radial magnetic
field within range of $\pm300$G. }
\end{figure*}

Compare to the previous case, the model M5 takes into account effects
of the non-axisymmetric field in the angular momentum balance via
the mean Maxwell stresses of the non-axisymmetric magnetic field,
i.e., terms like ${\displaystyle -\frac{\overline{\tilde{B}_{i}\tilde{B}_{j}}}{4\pi\overline{\rho}}}$
(over-bar means the azimuthal averaging) and contributions of the
non-axisymmetric field to the mean magnetic energy. The latter makes
effect to the efficiency of the magnetic quenching of the $\Lambda$
effect and coefficients of the eddy viscosity and thermal eddy conductivity.
Figure \ref{fig:M5e} shows evolutions of parameters of the large-scale
magnetic field and snapshot for the magnetic field distribution out
of the star. It is found that the non-axisymmetric dynamo quenches
generation of the axisymmetric magnetic field. However, unlike to
the model M3, which has the same parameter $C_{\alpha}$, the axisymmetric
dynamo persists in the stationary stage of evolution of the model
M5. This is similar to the model M2. The dynamo wave goes equator-ward
in the whole range of latitudes. The surface magnetic field is dominated
by the m=1 mode of the non-axisymmetric magnetic field with the Yin
Yang magnetic polarity pattern as well as the model M2.

\begin{figure}
\includegraphics[width=1\columnwidth]{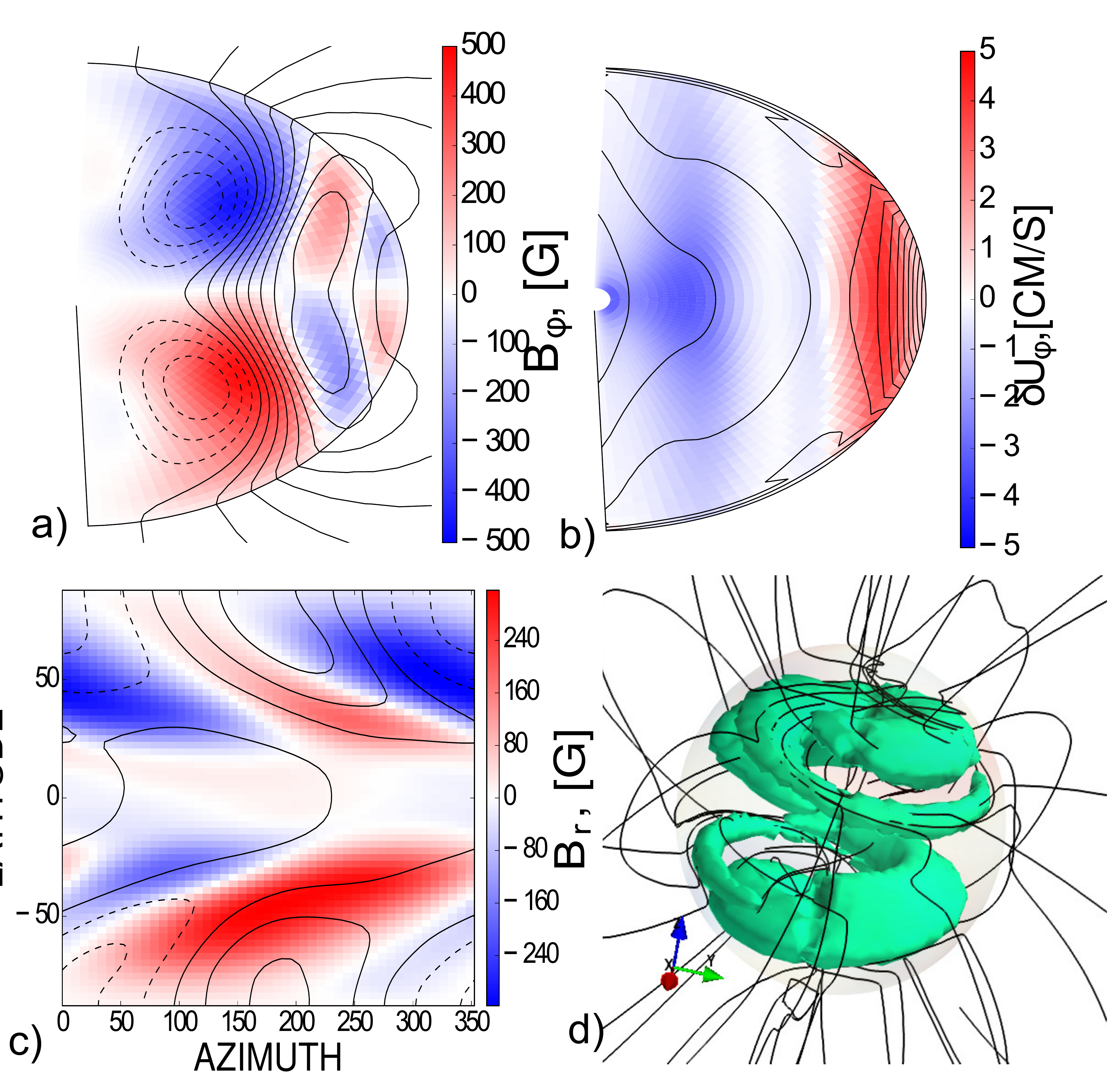}

\caption{\label{fig:M5_Snaps}Except the panel b) the same as Figure \ref{fig:M2}
for the model M5. The panel b) shows, the distribution of the mean
angular velocity (contours), and the background image shows distribution
of the zonal variations of rotation.}
\end{figure}

Figure \ref{fig:M5_Snaps} shows snapshots of the magnetic field and
angular velocity distributions in the star for the model M5. The snapshot
of the axisymmetric magnetic field is similar to those demonstrated
in the other models. The magnetic field is antisymmetric about equator
showing three bands of the toroidal magnetic field propagating outward
of the rotation axis, along the spherical iso-surfaces of the angular
velocity (see, Figure \ref{fig:M5_Snaps}b). The zonal variations
of the angular velocity are small and their patterns are elongated
along the axis of rotation. Similar to models M2 and M3, the large-scale
magnetic field inside the star has a spiral structure because the
non-axisymmetric mode m=1 dominates the others partial modes. 

In our simulations, we also have tried the larger values of the
parameters $Pm_{T}$ and $C_{\alpha}$.  Results for the  kinematic models with non-linear $\alpha$
effect and $Pm_{T}=10$ are similar  the model M3. The period of the axisymmetric dynamo in case of $Pm_{T}=10$ is about
120 years in agreement with expectations of \citet{shul15}. The model
show rather strong polar axisymmetric  magnetic field with the strength of
1.5kG. This is because of the  meridional circulation effect. Its
efficiency ncreases with  the increase of the parameter $Pm_T$. For
the $Pm_{T}=10$ the axisymmetric regime persists when the
$C_{\alpha}<0.04$. We found the transition
to the non-axisymmetric dynamo for the $C_{\alpha}=0.05$. The
properties  of the non-axisymmetric mean-field dynamo in the fully non-linear
regime for the case of  $Pm_T=10$  remain unclear.
% In whole, our results suggest that for this rotational period of 10 day the
% dynamo regime of the M-star show the regular large-scale non-axisymmetric
% field comes out robustly in the studied range of the governing parameters.

\section{Discussion}

The previous consideration of the mean-field models of the fully convective
stars was restricted to analysis of the eigenvalue problems \citep{elst07,shul15}
or the kinematic case with uniform density stratification and the
algebraic non-linearity of the $\alpha$-effect \citep{2006AA446.1027C}.
The main progress in theoretical understanding of the dynamo on the
the fully convective stars were made with help of the direct numerical
simulations (see, e.g., \citealt{2006ApJ...638..336D,2008ApJ676.1262B,2012AA546A19G,2015ApJ813L31Y}).
The paper for the first time presents results of the non-linear mean-field
dynamo models of the fully convective star rotating with period 10
days. 

The key reasons to study the mean-field models is to study behavior
of the dynamo in varying the governing dynamo parameters. At the first
step, let us discuss the kinematic dynamos with the nonlinear $\alpha$-effect.
The angular velocity profile in this case is different to the cylinder-like
pattern, which was discussed in the literature (see, e.g., \citealt{moss04,moss05,2006AA446.1027C})
and which appears in the direct numerical simulations. Our model include
effect of the meridional circulation which is important in the subsurface
layer for the case of $Pm_{T}>1$. For the case $Pm_{T}=3$, the model
M1 shows the strong axisymmetric dipole-like magnetic field with magnitude
of the polar field about 1kG. The dominance of the antisymmetric relative
to equator magnetic field disappears in the nonlinear mixed parity
solution if we neglect the meridional circulation. The dynamo waves
show the solar-like time-latitude diagrams with toroidal field drifting
to the equator and the radial field drifting to the pole.

\begin{figure}
\includegraphics[width=0.98\columnwidth]{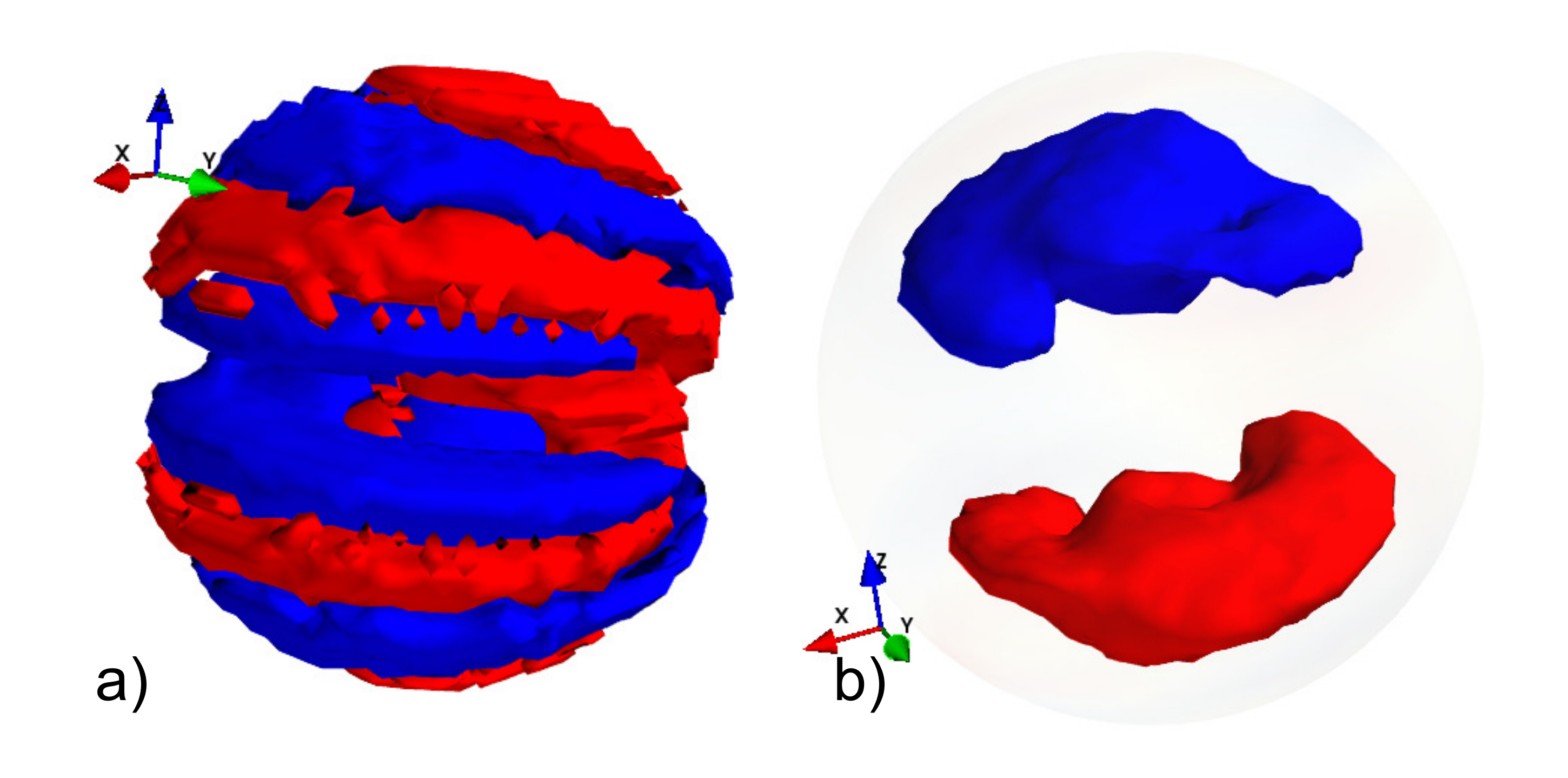}

\caption{\label{fig:alph}Snapshots of the model M5, a) the nonlinear $\alpha$
effects (volume contours for $\pm$3cm/s); b) the small-scale magnetic
helicity density (from the Eq(\ref{eq:helcon-1})), (volume contours
for $\pm$1.05$\cdot10^{10}$G$^{2}$/M)}

\end{figure}

The eigenvalue analysis shows that generation of the non-axisymmetric
magnetic field for the case of $Pm_{T}>1$ is less efficient than
the axisymmetric dynamo because the critical parameter of the dynamo
instability is smaller in the second case. This is general conclusion
of the most studies of the mean field dynamo starting from the seminal
paper by \citet{rad86AN}. The conclusion lead to ignorance of the
non-axisymmetric dynamos even for the super-critical regimes of the
axisymmetric dynamo (cf, \citealt{2016JFM799R6R}). However the model
M2 show that in case of the dynamo instability of the non-axisymmetric
field, the non-axisymmetric regime can beat the axisymmetric one.
The interaction between axisymmetric and non-axisymmetric magnetic
field goes via the nonlinear effects. Those are the conservation of
the magnetic helicity and the magnetic buoyancy. Contributions of
the magnetic helicity on the $\alpha$-effect can not be ignored in
the mean-field solar dynamos \citep{2007NJPh....9..305B}. They are
important in the non-axisymmetric dynamo, as well. The change of the
dynamo regime for the overcritical $C_{\alpha}$ is because of the
non-axisymmetric $\alpha$-effect, which is produced by the magnetic
helicity conservation in the non-axisymmetric large-scale dynamo.
Figures\ref{fig:alph}(a,b) show snapshots of the $\alpha_{\phi\phi}$
(see, Eq.(\ref{alp2d}) and the mean magnetic helicity density of
the small-scale field, which is generated because of the magnetic
helicity conservation in the model M5. Models M2 and M3 show similar
distributions. The models produce the non-axisymmetric non-linear
$\alpha$ effect and this supports dominance the non-axisymmetric
magnetic field in the dynamo. In the solar dynamo models the non-axisymmetric
$\alpha$-effect was employed for explanation of the so-called active
longitudes of the sunspot formations \citep{bigruz,berd06}. In our
models this effect stems naturally from magnetic helicity conservation.

The magnetic feedback on the differential rotation reduces efficiency
of the axisymmetric dynamo. The strength of the large-scale magnetic
field in the model M4 is less than in the model M1. The cyclic effect
of the large and small-scale Lorentz force on the angular momentum
fluxes produces phenomena known in the solar magnetic acitvity like
the zonal variations of the angular velocity and variations of the
meridional flow. Both of them predicted to have much smaller amplitude
than for the Sun. The rotational velocity at the equator is $1.44$
km/s, then the predicted magnitude of the latitudinal shear between
equator and pole is only about 14 m/s. Therefore our models demonstrate
the dynamo induced zonal variations are about of 10 percent of magnitude
of the mean latitudinal shear. The relative variations of the meridional
circulation are about 1 percent of the mean flow which is much smaller
than it is observed on the Sun. Note, the M-dwarf has much denser
plasma than the Sun and for the 1kG magnetic field at the top of the
integration domain ($0.98R_{\star}$) the Alfven velocity is less
than $6$m/s. In the model the toroidal field does not penetrate to
the surface because of the vacuum boundary conditions and this reduces
the magnitude of the large-scale flow variations on the surface. Unlike
the Sun (see, eg, \citealt{2011JPhCS.271a2001B,2011JPhCS271a2074H,2013ASPC..479..395K})
the predicted torsional oscillations have the equal magnitudes in
the bulk of the star and at the surface. Variations of the meridional
circulation are concentrated to the surface. Note , that the radial
profile of the meridional circulation is still unclear in the case
of the Sun, see preliminary results in the papers by \citet{hath12}
and \citet{Zhao13m}, who supports concentration of 11-th year variations
of the solar meridional circulation to the surface.

\begin{figure}
\includegraphics[width=1\columnwidth]{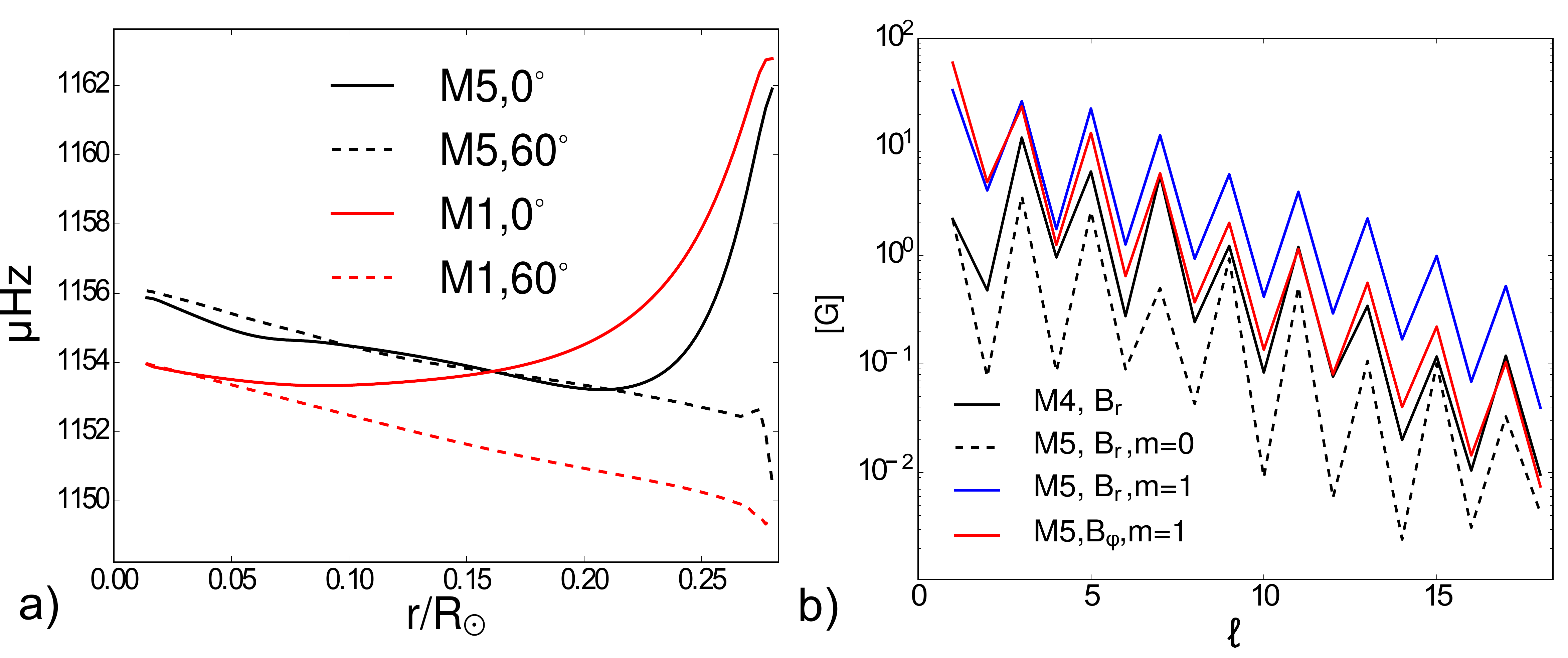}\caption{\label{fig:modes}a) The angular velocity radial profile in the kinematic
(red lines) and nonlinear models M1 and M5 for the equator ($0^{\circ}$)
and 60$^{\circ}$ latitudes; b)Modes.}
\end{figure}

It is predicted that magnetic activity produces rather strong distortion
of the angular velocity profile inside the star leaving the structure
of the meridional flow nearly the same as it is in the kinematic models.
The same results were found in the direct numerical simulations of
\citet{2008ApJ676.1262B} and \citet{2015ApJ813L31Y}. Figure \ref{fig:modes}a
allows comparison to their results. We find that in the magnetic case
(the model M5) the latitudinal shear persists only in the upper layer
of the star. Also, there the strong radial shear presents near the
equator. The same was found in the direct numerical simulation by
\citet{2015ApJ813L31Y}. The model of \citet{2008ApJ676.1262B} showed
the uniform angular velocity profile in the magnetic case. We find
that in the nonlinear model M5 the positive radial shear in the equatorial
region is stronger than in the kinematic model M1. Also we see formation
of the radial shear at the surface in the polar region in the model
M5. The increase of the magnitude of the subsurface shear as a result
of the magnetic field influence on the angular momentum fluxes is
also in agreement with the recent numerical simulations on the solar-like
stars (\citealt{guer2013,2014AA...570A..43K,2016ApJ819.104G}).

Our results show that the strength of the surface poloidal magnetic
field is only factor two or three lesser than the strength of the
toroidal magnetic field inside the star, see the Table 1. All the
models show rather strong polar magnetic field, 1kG in the kinematic
models and from 100 to 500 G in the nonlinear models. Current observations
of the stellar magnetic activity inform us a lot about the topological
and spectral properties of the magnetic field distributions at M-dwarfs
and cool stars \citep{M2-2010MNRAS,2016MNRAS1129S}. Figure \ref{fig:modes}b
presents results of the spherical harmonic decomposition for magnetic
field predicted by the fully nonlinear models M4 (axisymmetric one)
and M5. In the axisymmetric model M4 we don't expect any toroidal
field out of the surface because of the boundary conditions. In this
case the energy of the magnetic field outside the star is dominated
by $\ell=3$ and $\ell=5$ harmonics which is similar to the Sun \citep{sten88,2013AARv2166S,2016MNRAS4591533V}.
The non-axisymmetric dynamo model M5 show the dominance of the mode
m=1 and $\ell=1$ of the large-scale toroidal magnetic field. The
ratio of the energy of the non-axisymmetric and axisymmetric poloidal
magnetic field in the model M5 is about factor order of the magnitude.
The given results are in agreement with \citet{M2-2010MNRAS} for
the magnetic field observations for the early types of the M-dwarfs
with a moderate rotation rates.

Let's summarize the main findings of the paper. Our study confirm
the previous conclusions of \citet{shul15} that the weak differential
rotation of the M-dwarfs can support the axisymmetric dynamo especially
for the case $Pm_{T}>1.$ For the case $Pm_{T}=3$ we find that the
generation threshold $\alpha$-effect parameter $C_{\alpha}$ is lower
for axisymmetric magnetic field. However for the overcritical $\alpha$-effect
the non-axisymmetric dynamo become preferable. The situation is reproduced
both in the kinematic and in the fully nonlinear dynamo models. In
the non-linear case the differential rotation of the star deviates
strongly from the kinematic case. For the most complete non-linear
dynamo model we found the non-axisymmetric magnetic field of strength
about 0.5kG at at the surface mid latitude, it is rigidly rotating
and it is perturbed by the axisymmetric dynamo waves propagating out
of the rotational axis. The predicted dynamo period of the axisymmetric
dynamo waves in the model is about 40 Yr for the $Pm_T=3$ and it is
longer for the higher $Pm_T$.
\subsection*{Acknowledgements} 
I appreciate Prof D.D.~Sokoloff, Prof D.~Moss and Dr D.Shulyak for
discussions and comments. I thank the financial support of the project II.16.3.1
of ISTP SB RAS and the partial support of the  RFBR grants 15-02-01407-a, 16-52-50077-jaf. 
\bibliographystyle{mn2e}
%\bibliography{/home/va/work/pap/dyn}

\section{Appendix}

\setcounter{equation}{0}

\global\long\def\theequation{A\arabic{equation}}

\subsection{Heat transport\label{sub:Heat-transport}}

\citet{phd} found that under the joint action of the Coriolis force
and the large-scale\emph{ toroidal} magnetic field, and when it holds
$\Omega^{*}>1$, the eddy heat conductivity tensor could be approximated
as follows
\begin{equation}
\chi_{ij}\approx\chi_{T}\left(\phi_{\chi}^{(I)}\left(\beta\right)\phi\left(\Omega^{*}\right)\delta_{ij}+\phi_{\chi}^{(\|)}\left(\beta\right)\phi_{\parallel}\left(\Omega^{*}\right)\frac{\Omega_{i}\Omega_{j}}{\Omega^{2}}\right),\label{eq:ht-F}
\end{equation}
where 
\begin{eqnarray*}
\phi_{\chi}^{(I)} & = & \frac{2}{\beta^{2}}\left(1-\frac{1}{\sqrt{1+\beta^{2}}}\right),\\
\phi_{\chi}^{(\|)} & = & \frac{2}{\beta^{2}}\left(\sqrt{1+\beta^{2}}-1\right).
\end{eqnarray*}
Expression \ref{eq:ht-F} were obtained the standard schemes of the
mean-field magnetohydrodynamics employing the so-called ``second
order correlation approximation'' and the mixing length approximations.
Also we skip components of the tensor along the large-scale magnetic
field.

The heat transport by radiation reads, 
\[
\mathbf{F}^{rad}=-c_{p}\overline{\rho}\chi_{D}\boldsymbol{\nabla}T,
\]
where 
\[
\chi_{D}=\frac{16\sigma\overline{T}^{3}}{3\kappa\overline{\rho}^{2}c_{p}},
\]
where $\kappa$ is opacity coefficient. The radial profiles of the
gravity acceleration, $g$, the density, $\overline{\rho}$, the temperature,
$\overline{T}$, the heat source, $\epsilon$, as well as others thermodynamic
parameters, like the $c_{p}$ or the $\kappa$ are taken form the
reference model which was calculated with help of the MESA code.

\subsection{Angular momentum balance}

Expression of the Reynolds is determined from the mean-field hydrodynamics
theory (see, \citealt{1994AN....315..157K,kit2004AR}) as follows
\begin{equation}
\hat{T}_{ij}=\left(\left\langle u_{i}u_{j}\right\rangle -\frac{1}{4\pi\overline{\rho}}\left(\left\langle b_{i}b_{j}\right\rangle -\frac{1}{2}\delta_{ij}\left\langle \mathbf{b}^{2}\right\rangle \right)\right),\label{eq:stres}
\end{equation}
where $\mathbf{u}$ and $\mathbf{b}$ are fluctuating velocity and
magnetic fields. The turbulent stresses take into account the turbulent
viscosity and generation of the large-scale shear due to the $\Lambda$-
effect \citep{kit11}: 
\begin{eqnarray}
T_{r\phi} & = & \overline{\rho}\nu_{T}\left\{ \Phi_{\perp}+\left(\Phi_{\|}-\Phi_{\perp}\right)\mu^{2}\right\} r\frac{\partial\sin\theta\Omega}{\partial r}\nonumber \\
 & + & \overline{\rho}\nu_{T}\sin\theta\left(\Phi_{\|}-\Phi_{\perp}\right)\left(1-\mu^{2}\right)\frac{\partial\Omega}{\partial\mu}\label{eq:trf}\\
 & - & \overline{\rho}\nu_{T}\sin\theta\Omega\left(\frac{\alpha_{MLT}}{\gamma}\right)^{2}\left(V^{(0)}+\sin^{2}\theta V^{(1)}\right),\nonumber \\
T_{\theta\phi} & = & \overline{\rho}\nu_{T}\sin^{2}\theta\left\{ \Phi_{\perp}+\left(\Phi_{\|}-\Phi_{\perp}\right)\sin^{2}\theta\right\} \frac{\partial\Omega}{\partial\mu}\nonumber \\
 & + & \overline{\rho}\nu_{T}\left(\Phi_{\|}-\Phi_{\perp}\right)\mu\sin^{2}\theta r\frac{\partial\Omega}{\partial r}\label{eq:ttf}\\
 & + & \overline{\rho}\nu_{T}\mu\Omega\sin^{4}\theta\left(\frac{\alpha_{MLT}}{\gamma}\right)^{2}H^{(1)},\nonumber 
\end{eqnarray}
where $\nu_{T}={\displaystyle \frac{4}{5}\eta_{T}}$. The viscosity
functions - $\Phi_{\|},\Phi_{\perp}$ and the $\Lambda$- effect -
$V^{\left(0,1\right)}$ and $H^{\left(1\right)}$, are dependent on
the Coriolis number and the strength of the large-scale magnetic field.
They also depends on the anisotropy of the convective flows. Similar
to the Subsection\ref{sub:Heat-transport} we employ the fast rotating
regime of the magnetic quenching for the eddy viscosity and the the
$\Lambda$- effect as it was discussed earlier in \citep{kuetal96,p99,phd}:
\begin{eqnarray}
\Phi_{\perp}\!\!\! & =\!\!\! & \!\!\psi_{\perp}\left(\Omega^{\star}\right)\phi_{V\perp}\left(\beta\right),\,\,\Phi_{\parallel}=\psi_{\parallel}\left(\Omega^{\star}\right)\phi_{\chi}^{(I)}\left(\beta\right),\label{visc-f}\\
V^{(0)}\!\!\! & =\!\!\! & \!\!\left(J_{0}\left(\Omega^{\star}\right)\!+\!J_{1}\left(\Omega^{\star}\right)\!+\!a\left(I_{0}\left(\Omega^{\star}\right)\!+\!I_{1}\!\left(\Omega^{\star}\right)\!\!\right)\!\right)\!\phi_{\chi}^{(I)}\!\!\left(\beta\right)\!,\label{v0-f}\\
V^{(1)}\!\!\! & =\!\!\! & \!\!\left(J_{1}\left(\Omega^{\star}\right)+aI_{1}\left(\Omega^{\star}\right)\right)\phi_{\chi}^{(I)}\left(\beta\right),\label{v1-f}\\
H^{(0)}\!\!\! & =\!\!\! & \!\!J_{4}\left(\Omega^{\star}\right)\phi_{H}\left(\beta\right),\label{h0-f}
\end{eqnarray}
and $H^{(1)}=-V^{(1)}$, where the new magnetic quenching functions
are:
\begin{eqnarray}
\phi_{V\perp} & = & \frac{4}{\beta^{4}\sqrt{\left(1+\beta^{2}\right)^{3}}}\left(\left(\beta^{4}+19\beta^{2}+18\right)\sqrt{\left(1+\beta^{2}\right)}\right.\nonumber \\
 & - & \left.8\beta^{4}-28\beta^{2}-18\right),\\
\phi_{H} & = & \frac{4}{\beta^{2}}\left(\frac{2+3\beta^{2}}{2\sqrt{\left(1+\beta^{2}\right)^{3}}}-1\right).
\end{eqnarray}
We employ the parameter of the turbulence anisotropy $a=1$ (see discussion,
by \citealt{kit2004AR}). The equation \ref{visc-f}) shows that for
case of the ``fast'' rotating fluid the large-scale magnetic field
quenches the eddy viscosity anisotropy. This conclusion was obtained
for the case when the toroidal large-scale magnetic field dominates
the poloidal component \citep{phd}. This approximation may be incorrect
for the fast rotating fully convective stars. 

The Lambda effect is modulated by the factor $\ell\left|\boldsymbol{\Lambda}^{(\rho)}\right|{\displaystyle \approx\frac{\alpha_{MLT}}{\gamma}}$,
where $\boldsymbol{\Lambda}^{(\rho)}=\boldsymbol{\nabla}\log\overline{\rho}$.
It varies sharply near the center and the top of the star. To avoid
the numerical complications we force the $\Lambda$-effect to go zero
toward the center of the star, we replaced that factor as follows,
\begin{equation}
{\displaystyle \frac{\alpha_{MLT}}{\gamma}=\frac{\alpha_{MLT}}{2\gamma}\left(1+erf\left(50\left(\frac{r}{R_{\star}}-.1\right)\right)\right)},\label{eq:mlt}
\end{equation}
where $\alpha_{MLT}=1.91$ and $\gamma={\displaystyle \frac{5}{3}}$. 

The first term in the RHS of the Eq.(\ref{eq:vort}) describes dissipation
of the mean vorticity, $\omega$. Similarly to \citet{rem2005ApJ}
we approximate it as follows, 
\begin{equation}
-\left[\boldsymbol{\nabla}\times\frac{1}{\overline{\rho}}\boldsymbol{\nabla\cdot}\overline{\rho}\hat{\mathbf{T}}\right]_{\phi}\approx2\nu_{T}\phi_{1}\left(\Omega^{*}\right)\psi_{1}\left(\beta\right)\nabla^{2}\omega,\label{eq:vort2}
\end{equation}
where $\nu_{T}={\displaystyle \frac{4}{5}\chi_{T}},$ the rotational
function $\phi_{1}$ and the magnetic quenching function are given
in \citet{1994AN....315..157K}. We have tried the more general formalism
with full components of the eddy-viscosity tensor for the rotating
turbulence provided by \citet{1994AN....315..157K}. We found results
to be similar to the case of the Eq(\ref{eq:vort2}). 

For the ideal gas the last term in Eq.(\ref{eq:vort}) can be rewritten
in terms of the specific entropy \citep{kit11}, 
\begin{equation}
\frac{1}{\overline{\rho}^{2}}\left[\boldsymbol{\nabla}\overline{\rho}\times\boldsymbol{\nabla}\overline{p}\right]_{\phi}\approx-\frac{g}{rc_{p}}\frac{\partial s}{\partial\theta}.\label{eq:baroc}
\end{equation}
The meridional circulation is expressed by a stream function $\Psi$,
$\overline{\mathbf{U}}^{m}={\displaystyle \frac{1}{\overline{\rho}}}\boldsymbol{\nabla}\times\hat{\boldsymbol{\phi}}\Psi$.
The $\Psi$ and the $\omega$ are related via the equation 
\begin{equation}
-\overline{\rho}\omega=\left(\Delta-\frac{1}{r^{2}\sin^{2}\theta}\right)\Psi-\frac{1}{r\overline{\rho}}\frac{\partial\overline{\rho}}{\partial r}\frac{\partial r\Psi}{\partial r}.\label{eq:poiss}
\end{equation}

We employ the stress-free boundary conditions for the Eq.(\ref{eq:az}),
the azimuthal component of the mean vorticity, $\omega$, is put to
zero at the boundaries.

\subsection{The mean-electromotive force}

This section of Appendix describe some parts of the mean-electromotive
force. The basic formulation is given in (Pipin, 2008) (hereafter,
P08). For this paper we reformulate tensor $\alpha_{i,j}^{(H)}$,
which represents the hydrodynamical part of the $\alpha$-effect,
by using Eq.(23) from P08 in the following form, 
\begin{eqnarray}
\alpha_{ij}^{(H)} & = & 3\frac{\left(\mathbf{\boldsymbol{\Omega}}\cdot\boldsymbol{\Lambda}^{(\rho)}\right)}{\Omega}\left\{ \delta_{ij}f_{10}^{(a)}+\frac{\Omega_{i}\Omega_{j}}{\Omega^{2}}f_{5}^{(a)}\right\} ,\label{alfh}
\end{eqnarray}
where $\boldsymbol{\Lambda}^{(\rho)}=\boldsymbol{\nabla}\log\overline{\rho}$
. The other parts of the $\alpha$-effect are rather small because
the star is in the regime of the fast rotation, when the Coriolis
number $\Omega^{*}\gg1$. Moreover, if we neglect terms order of $O\left({\displaystyle \frac{1}{\Omega^{*}}}\right)$
in the Taylor expansion of the Eq.(\ref{alfh}), we get (\citep{kit-rud:1993b}):
\[
\alpha_{ij}^{(H)}=-\frac{3\pi}{2}\frac{\left(\mathbf{\boldsymbol{\Omega}}\cdot\boldsymbol{\Lambda}^{(\rho)}\right)}{\Omega}\left\{ \delta_{ij}-\frac{\Omega_{i}\Omega_{j}}{\Omega^{2}}\right\} ,
\]
The functions $f_{5,10}^{(a)}$ where defined in P08 for the general
case which includes the effects the hydrodynamic and magnetic fluctuations
in the background turbulence. In the paper we employ the case when
the background turbulent fluctuations of the small-scale magnetic
field are in the equipartition with the hydrodynamic fluctuations,
i.e., ${\displaystyle \varepsilon=\frac{b'^{2}}{4\pi\overline{\rho}u'^{2}}}=1$,
where the $u'^{2}$ and $b'^{2}$ are intensity of the background
turbulent velocity and magnetic field. The magnetic quenching function
of the hydrodynamical part of $\alpha$-effect is defined by 
\begin{equation}
\psi_{\alpha}=\frac{5}{128\beta^{4}}\left(16\beta^{2}-3-3\left(4\beta^{2}-1\right)\frac{\arctan\left(2\beta\right)}{2\beta}\right),
\end{equation}
The magnetic helicity part of the $\alpha$-effect, $\alpha_{i,j}^{(M)}$
is expressed by 
\begin{eqnarray}
\alpha_{ij}^{(M)} & = & \left\{ \delta_{ij}f_{2}^{(a)}\left(\Omega^{*}\right)-\frac{\Omega_{i}\Omega_{j}}{\Omega^{2}}f_{1}^{(a)}\left(\Omega^{*}\right)\right\} .\label{alfm}
\end{eqnarray}
We employ the anisotropic diffusion tensor which is derived in P08
and in \citep{2014ApJ_pipk}: 
\begin{eqnarray}
\eta_{ijk} & = & 3\eta_{T}\left\{ \left(2f_{1}^{(a)}-f_{2}^{(d)}\right)\varepsilon_{ijk}+2f_{1}^{(a)}\frac{\Omega_{i}\Omega_{n}}{\Omega^{2}}\varepsilon_{jnk}\right\} \label{eq:diff}\\
 & + & a\eta_{T}\phi_{1}\left(g_{n}g_{j}\varepsilon_{ink}-\varepsilon_{ijk}\right)\nonumber 
\end{eqnarray}
where $\mathbf{g}$ is the unit vector in the radial direction, $a=1$
is the parameter of the turbulence anisotropy, $\eta_{T}$ is the
magnetic diffusion coefficient. The quenching functions $f_{1,2}^{(a,d)}$and
$\phi_{1}$ are given in \citep{2014ApJ_pipk}.

The turbulent pumping of the mean-field contains the sum of the contributions
due to the mean density gradient \citep{kit:1991} and the mean-filed
magnetic buoyancy \citep{kp93}, $\gamma_{ij}^{(b)}$,:

\begin{equation}
\gamma_{ij}=\gamma_{ij}^{(\rho)}+\gamma_{ij}^{(b)},\label{eq:pump}
\end{equation}
where each contribution is defined as follows:

\begin{eqnarray}
\gamma_{ij}^{(\rho)} & = & 3\eta_{T}\left\{ f_{3}^{(a)}\Lambda_{n}^{(\rho)}+f_{1}^{(a)}\left(\mathbf{e}\cdot\boldsymbol{\Lambda}^{(\rho)}\right)e_{n}\right\} \varepsilon_{inj}\label{eq:pumpr}\\
 &  & -3\eta_{T}f_{1}^{(a)}e_{j}\varepsilon_{inm}e_{n}\Lambda_{m}^{(\rho)}\nonumber \\
\gamma_{ij}^{(b)} & = & \frac{\alpha_{MLT}u'}{\gamma}\beta^{2}K\left(\beta\right)g_{n}\varepsilon_{inj},\label{eq:pumpb}
\end{eqnarray}
where $f_{1,3}^{(a)}\left(\Omega^{\star}\right)$ are functions of
the Coriolis number, $u'$ is the RMS of the convective velocity,
$\mathbf{\boldsymbol{\Lambda}}_{i}^{(\rho)}=\boldsymbol{\nabla}_{i}\log\overline{\rho}$
are components of the gradient of the mean density. The $\alpha_{MLT}$
is the parameter of the mixing length theory, $\gamma$ is the adiabatic
exponent and the function $K\left(\beta\right)$ is defined in \citep{kp93}
and $\mathbf{g}$ is the unit vector in the radial direction. 
\end{document}